\documentclass{jfm}
\usepackage{graphicx}
\usepackage{epstopdf, epsfig}
\usepackage{csquotes}
\usepackage{enumitem}

\newcommand{\bo}[1]{\boldsymbol{#1}}

\def\XXint#1#2#3{{\setbox0=\hbox{$#1{#2#3}{\int}$}
     \vcenter{\hbox{$#2#3$}}\kern-.5\wd0}}

\shortauthor{S. Nambiar, P. Garg and G. Subramanian}

\title{Enhanced velocity fluctuations in interacting swimmer suspensions}

\author{Sankalp Nambiar,
Piyush Garg
 \and Ganesh Subramanian   \corresp{\email{sganesh@jncasr.ac.in}}}

\affiliation{Engineering Mechanics Unit, Jawaharlal Nehru Centre for Advanced Scientific Research, Jakkuru, Bangalore, 560064, India}

\begin{document}

\maketitle

\begin{abstract}
This paper presents an analytical characterization of the fluid velocity fluctuations in dilute suspensions of hydrodynamically interacting slender micro-swimmers. The velocity variance is \textit{O}($nL^3$) and finite for a non-interacting suspension, with the covariance decaying as \textit{O}($1/r$) on scales larger than $L$; here, $nL^3 <$ \textit{O}(1) is the swimmer (hydrodynamic) volume fraction, with $n$ being the swimmer number density and $L$ its characteristic length. For a suspension of interacting straight-swimmers, however, pair-correlations result in a non-decaying velocity covariance, with a variance that, at \textit{O}$(nL^3)^2$, diverges logarithmically with system size, consistent with the results of earlier numerical simulations \citep{underhill2011}. This latter divergence is arrested on the inclusion of an orientation decorrelation mechanism - either rotary diffusion or run-and-tumble dynamics. Dilute suspensions of hydrodynamically interacting run-and-tumble particles (RTPs) are examined in detail as a function of the dimensionless run length $U\tau/L$; here, $U$ is the isolated swimmer swimming speed and $\tau$ its mean run duration. The velocity variance, at \textit{O}$(nL^3)^2$, transitions from an initial linear increase for $U\tau/L \ll 1$, to an eventual logarithmic increase for $U\tau/L \gg 1$, the latter being consistent with the divergence in the straight-swimmer limit ($U\tau/L \rightarrow \infty$). Suspensions of interacting pushers exhibit a greater velocity variance for all $U\tau/L$. The mean square displacement of immersed passive tracers exhibits an increasingly broad crossover from the ballistic to the diffusive regime, for large $U\tau/L$, on account of swimmer interactions, with the tracer diffusivity at \textit{O}$(nL^3)^2$ scaling as \textit{O}($U\tau/L$) for $U\tau/L \gg 1$. Our analysis explains numerous observations of a volume-fraction-dependent crossover time for the passive mean square displacement, and the bifurcation of the velocity variance and tracer diffusivities between pusher and puller suspensions.

\end{abstract}


\section{Introduction}\label{sec:intro}
Suspensions of rear-actuated swimming microorganisms (pushers), such as bacteria \emph{E. coli} and \emph{B. subtilis}, exhibit a state of large-scale coherent motion that arises, in part, due to long-ranged hydrodynamic interactions \citep{simha02, toner2005hydrodynamics, saintillan07, saintillan2008, G2008, subkoch2009, ramaswamy2010mechanics, subprabhu2011, koch2011, underhill2011, marchetti2013, deepak2015, clement16}. Experiments \citep{dombrowski2004, clement2014} and simulations \citep{saintillan07, yoemans2012, deepak2015} have shown that the transition to collective dynamics occurs beyond a threshold concentration, leading to \enquote*{bacterial turbulence} \citep{yoemans2012}; very recent experiments, in fact, point to the dominant role of hydrodynamic interactions in this regard \citep{drescher2019}. Collective motion has important consequences for both transport and rheology, with experiments and mean-field theories having shown a reduction in viscosity leading to apparent superfluidity and unexpected shear-banding behavior \citep{Sokolov2009, Saintillan2010, gachelin2015, nambiar17, saintillan2018, Saintillan2010, nambiar2018, laxman2018, samanta18}. This transition has often been inferred from an anomalous enhancement in the diffusivities of passive tracer particles \citep{Wu2000, yodh2007, G2008,thiffeault2011, poon13, kasyapkoch2014, thiffeault2015, deepak2015, stenhammar2017, morozov2019}.

A large body of theoretical work studying swimmer suspensions relies on phenomenological and mean-field models \citep{saintillan2008, G2008, subkoch2009, simha02, aranson2007, yoemans2012, goldstein2013, dunkel2013b, marchetti2013, dunkel2016}. For a dilute swimmer suspension, mean-field theory has shown that it is the long-ranged hydrodynamic interactions \citep{G2008, saintillan2008, subkoch2009} between pushers and the mutually reinforcing orientation and velocity fluctuations that are responsible for the said transition to collective motion. In this paper, we go beyond this limiting mean-field assumption, and demonstrate the crucial role played by hydrodynamically induced swimmer-correlations. The interplay of swimming with long-ranged hydrodynamic interactions is shown, for the first time, to lead to a larger velocity variance, and thence, larger tracer diffusivities, in suspensions of run-and-tumble pushers vis-a-vis pullers (the swimming mechanism for pullers is front-actuated, as is typically the case for algae; see \cite{kessler1986individual, saintillan07, lauga2009, Saintillan2010, koch2011, guasto2012fluid, deepak2015, goldstein2015green}). Interactions also lead to a logarithmic divergence of the velocity variance in the straight-swimmer limit, that has been observed in earlier simulations \citep{underhill2011}.  The prediction of larger velocity fluctuations in pusher suspensions is also consistent with more recent simulations which have revealed a bifurcation of the fluid velocity variance in pusher and puller suspensions beyond a threshold \citep{deepak2015, stenhammar2017}.


\begin{figure}
\includegraphics[width=\textwidth]{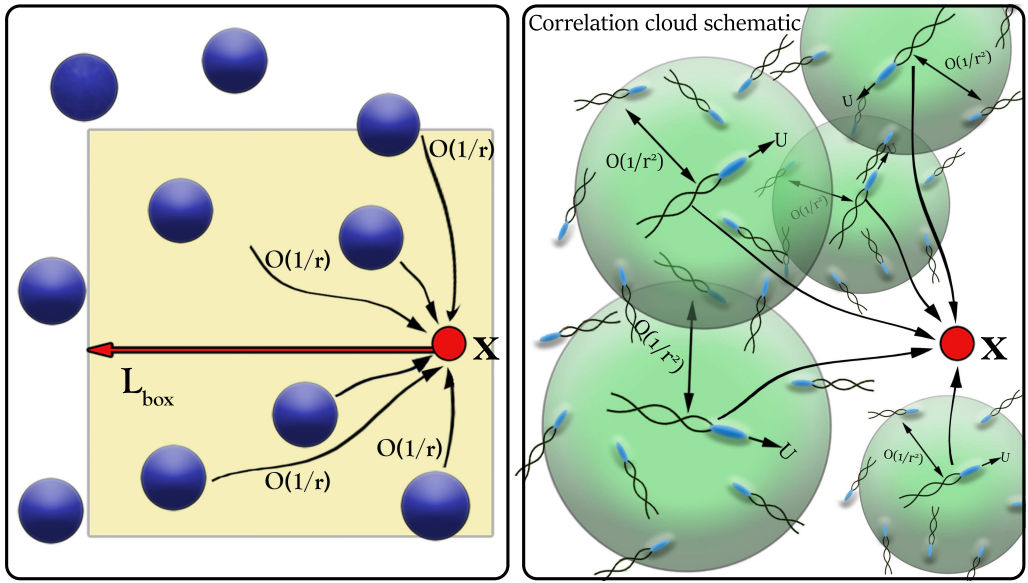}
\caption{\label{fig:schematic}The window on the left shows the \textit{O}($1/r$) contributions due to individual spheres contributing to the variance at $\boldsymbol{x}$ in an (infinite) sedimenting suspension. The right window corresponds to a straight-swimmer suspension with each swimmer surrounded by a pair-correlation cloud wherein correlations decay as \textit{O}($1/r^2$). The variance at $\boldsymbol{x}$ is the sum of the contributions of these individual clouds, and is logarithmically divergent.}
\end{figure}

In a homogeneous Stokesian suspension of sedimenting particles with a random microstructure, the velocity variance is predicted to diverge linearly with system size \citep{calfisch1985, hinch1988hydrodynamics, koch1991, nicolai1995, ladd1996, segre1997, ladd1997, sriram1998, ramaswamy2001, guazzelli2011, witten2017}. The resolution of this divergence has been a long-standing theoretical challenge \citep{hinch1988hydrodynamics, ramaswamy2001, guazzelli2011}. It arises due to the long-ranged \textit{O}($1/r$) disturbance velocity fields of the individual particles acting as point-forces (monopoles), as shown in figure \ref{fig:schematic}a. The fluid velocity variance at a given point $\boldsymbol{x}$ may be estimated as $\langle\boldsymbol{u}(\boldsymbol{x})\cdot\boldsymbol{u}(\boldsymbol{x})\rangle \equiv  \int nL^3U_s^2(1/r)^2 \mathrm{d}\boldsymbol{r} \sim (nL^3) U_s^2 L_{box}/L$. Here, $U_s$ is the mean sedimenting speed, $L_{box}$ the system size and $nL^3$ the hydrodynamic volume fraction, with $n$ being the particle number density, and $L$ a characteristic particle size. Unlike passive particles, microswimmers, both pushers and pullers, act as force-dipoles in the far-field, leading to a disturbance velocity field that decays more rapidly as \textit{O}($1/r^2$). An argument along lines similar to that for the passive Stokesian suspension above gives $\langle\boldsymbol{u}(\boldsymbol{x})\cdot\boldsymbol{u}(\boldsymbol{x})\rangle \equiv \int nL^3U^2(1/r)^4 \mathrm{d}\boldsymbol{r} \sim (nL^3) U^2$, $U$ being the swimming speed, implying that the velocity variance remains finite in the limit $nL^3 \ll 1$ with correlations between swimmers being neglected at leading order (the divergence at small $r$ implied in the scaling integral above is regularized once the finite swimmer size is accounted for); the covariance $\langle\boldsymbol{u}(\boldsymbol{x})\cdot\boldsymbol{u}(\boldsymbol{x}+\boldsymbol{r})\rangle \sim (nL^3) \int U^2 /(r-r^\prime)^4 \mathrm{d}\boldsymbol{r}^\prime$ at this order exhibits an \textit{O}($1/r$) decay for $r\gg L$ \citep{G2008, underhill2011}. Introducing pair-level correlations in straight-swimmer suspensions, as shown in figure \ref{fig:schematic}b, leads to a pair-orientation probability density that decays as \textit{O}($1/r^2$) in the far-field. In this scenario, where a given swimmer interacts pair-wise with a cloud of swimmers surrounding it, one has the fluid velocity variance due to \enquote*{each cloud} scaling as $[\boldsymbol{u}(\boldsymbol{x})\cdot\boldsymbol{u}(\boldsymbol{x})]_{cloud} \equiv nL^3\int U^2 r^{\prime 2}\mathrm{d}r^\prime/(r^4 r^{\prime 2}) \sim nL^3 U^2/r^3$, which when integrated over all such correlation clouds gives $\langle\boldsymbol{u}(\boldsymbol{x})\cdot\boldsymbol{u}(\boldsymbol{x})\rangle \equiv (nL^3)^2 \ln(L_{box}/L)$; a logarithmically divergent variance. For passive suspensions, early theoretical analyses have predicted a Debye-like screening of the long-ranged hydrodynamic interactions at distances of \textit{O}$(nL^2)^{-1}$ \citep{koch1991}, leading to a finite variance. In contrast, the divergence in experiments is cut off due to the development of a container-scale stratification \citep{luke2000, ramaswamy2001, guazzelli2011}. Clearly, the fact that correlations in active suspensions (of straight-swimmers) act to yield a divergent variance is in sharp contrast to the nature of velocity fluctuations in passive suspensions, and highlights the novel consequences of activity (swimming).


The manuscript is broadly organized into three sections. In \textsection\ref{sec:fcovar}, we examine the fluid velocity covariance. Following the formulation of the general expression, in \textsection\ref{subsec:noint}, we determine the covariance for suspensions of non-interacting swimmers, in which case the covariance is independent of the mechanism by which the swimmer orientation decorrelates. Next, in \textsection\ref{subsec:int}, we examine the covariance for the more involved case of a dilute suspension of interacting swimmers. Here, it is necessary to discuss the case of straight-swimmers and run-and-tumble particles (RTPs) separately, and this is done in \textsection\ref{subsubsec:intSt} and \textsection\ref{subsubsec:RTP}, respectively. The calculation of the correlated contribution, at \textit{O}$(nL^3)^2$,  requires the steady state pair probability density in position-orientation (phase) space, for both straight-swimmers and RTPs, which in turn requires obtaining an expression for the rotation rate experienced by a slender swimmer, due to the disturbance velocity field generated by another; the latter is derived in appendix \ref{sec:appRotationRate}. For straight swimmers, the phase space probability density is directly obtained in terms of generalized functions. For RTPs, the pair-probability density is again obtained in closed form, now as a function of the dimensionless parameter $U\tau/L$ which measures the run length in units of the swimmer size. Straight-swimmers correspond to the  (singular) limit $U\tau/L \to \infty.$ The resulting expression for the correlated component of the variance is shown to scale linearly with $U\tau/L$ in the limit $U\tau/L \ll 1$, while increasing logarithmically in the limit $U\tau/L \gg 1$, corresponding to persistent swimmers.  For $U\tau/L \ll 1$, the covariance in suspensions of RTPs transitions directly from the variance plateau for $r \ll 1$ to an \textit{O}($1/r$) far-field decay for $r \gg 1$. However, for large $U\tau/L$, we observe a weak decay of the covariance in the interval $1 \ll r \ll U\tau/L$ that is intermediate between the variance plateau ($r \ll 1$) and the aforementioned \textit{O}($1/r$) far-field decay regime ($r \gg U\tau/L$); the covariance remains non-decaying in the singular straight-swimmer limit. For RTPs in the limit $U\tau/L \gg 1$, we present an alternate derivation of both the variance and the covariance using a matched asymptotic expansions approach in \textsection\ref{subsubsec:matched_asymptotic}, which is shown to compare well with the exact result for $U\tau/L \geq$ \textit{O}(1). Importantly, this approach shows that the aforementioned weak intermediate decay of the covariance for $U\tau/L\gg 1$ is a logarithmic one. Next, in \textsection\ref{sec:tracers}, we study the diffusivity of immersed passive tracers. Here too, we first consider suspensions of non-interacting swimmers in \textsection\ref{subsec:msd_sing}, and then interacting swimmers in \textsection\ref{subsec:msd_int}. Finally, in \textsection\ref{sec:conclusion}, we present concluding remarks and a course for future work. We also briefly discuss the nature of pair-orientation correlations in Appendix \ref{sec:orientation}.

\section{The fluid velocity covariance}\label{sec:fcovar}
We begin with a discrete formulation applicable to a suspension of $N$ slender swimmers, each having a length $L$ and swimming with speed $U$ directed along its axis. The configuration of the swimmer suspension is characterized by the positions ($\boldsymbol{x}_i$) and orientations ($\boldsymbol{p}_i$) of the $N$ swimmers, with the swimmer number density field being defined as $c(\boldsymbol{x},t) = \sum_{i=1}^N \boldsymbol{\delta}(\boldsymbol{x}-\boldsymbol{x}_i)$. The disturbance velocity and pressure fields $\boldsymbol{u}_i$ and $P_i$, due to the $i^{th}$ swimmer, at leading logarithmic order, satisfy the Stokes equations forced by a line distribution of Stokeslets along the swimmer axis, and the equation of continuity:
\begin{subeqnarray}
-\bnabla P_i + \eta \nabla^2 \boldsymbol{u}_i(\boldsymbol{x}) &=& \int_{-L/2}^{L/2} f(s) \boldsymbol{p}_i \boldsymbol{\delta}(\boldsymbol{x} -\boldsymbol{x}_i - s\boldsymbol{p}_i) \mathrm{d}s, \label{stokes}\\[3pt]
	\bnabla\cdot\boldsymbol{u}_i &=& 0. \label{eq:continuity}
\end{subeqnarray}
Here, $\eta$ is the viscosity of the suspending fluid and $\boldsymbol{\delta}(\bo{z})$ represents the Dirac-delta function. Assuming the swimmers to be fore-aft symmetric and force-free, the linear force density along the swimmer axis (the axial coordinate being $s$)  may be expressed as $f(s) = \eta U sgn(s)/\ln\kappa$. For non-fore-aft symmetric swimmers, there is an \textit{O}(1) change in the force density, and thence the disturbance fluid velocity it generates, however, none of the principal conclusions detailed below change on account of swimmer asymmetry. The velocity disturbance satisfying (\ref{stokes}) may be written in terms of the Oseen-Burgers tensor $\mathsfbi{G}(\boldsymbol{x}) = 1/(8\pi\eta r)[\mathsfbi{I} + \bo{xx}/r^2]$ as:
\begin{equation}
\boldsymbol{u}_i(\boldsymbol{x}, t) = \int_{-L/2}^{L/2} f(s) \, \mathsfbi{G}(\boldsymbol{x}-\boldsymbol{x}_i - s\boldsymbol{p}_i)\bo{\cdot} \boldsymbol{p}_i \boldsymbol{d}s,
\label{stokes:oseens}
\end{equation}
and therefore, the suspension velocity field is expressible as:
\begin{equation}
\bo{u}(\bo{x}, t) \equiv \sum_{i=1}^N \bo{u}_i(\bo{x}, t) = \sum_{i=1}^N\int_{-L/2}^{L/2} f(s) \, \mathsfbi{G}(\bo{x}-\bo{x}_i - s\bo{p}_i)\bo{\cdot} \bo{p}_i \bo{d}s,
\label{eq:Suspensionoseens}
\end{equation}
where the time dependence comes from the evolving swimmer positions and orientations. Now, the fluid velocity covariance in a swimmer suspension is defined as:
\begin{eqnarray}
 \langle\boldsymbol{u}(\boldsymbol{x}, t)\bo{\cdot} \boldsymbol{u}(\boldsymbol{x}^\prime, t)\rangle & = & \Bigg\langle\sum_{i=1}^N\sum_{j=1}^N  \int_{-L/2}^{L/2} f(s) \mathrm{d}s  \int_{-L/2}^{L/2}  f(s^\prime)\mathrm{d}s^\prime \bo{\delta}(\boldsymbol{x}-\boldsymbol{x}_i)\bo{\delta}(\boldsymbol{p}-\boldsymbol{p}_i) \delta(\boldsymbol{p}^\prime-\boldsymbol{p}_j) \nonumber\\
&& \mbox{}\bo{\delta}(\boldsymbol{x}^\prime-\boldsymbol{x}_j)[\mathsfbi{G}(\boldsymbol{x}-\boldsymbol{x}_i - s\boldsymbol{p}_i)\bo{\cdot} \boldsymbol{p}_i]\bo{\cdot}[\mathsfbi{G}(\boldsymbol{x}^\prime-\boldsymbol{x}_j - s\boldsymbol{p}_j)\bo{\cdot} \boldsymbol{p}_j]\Bigg\rangle,
\label{covar1}
\end{eqnarray}
where the angular brackets denote an ensemble average ($\langle\boldsymbol{\cdot}\rangle$) over the configurations of swimmers $i, j$. Since the focus in this section is on the single-time covariance, we will neglect mention of the time dependence from hereon. Within the continuum framework ($N\gg1$), the above average is expressible in terms of the configurational (position-orientation) probability density function. To see this, one may split the double summation in (\ref{covar1}) into two distinct contributions; one containing only the diagonal terms, with $i = j$, which involves the singlet probability density defined as $\Omega_ 1 = \langle\sum_{i=1}^N \delta(\boldsymbol{x}_1-\boldsymbol{x}_i)\delta(\boldsymbol{p}_1-\boldsymbol{p}_i)\rangle$; and a second contribution containing the off-diagonal terms $i\neq j$ that involves the pair probability density defined as $\Omega_2$ = $\langle\sum_{i, j=1; i\neq j}^N\bo{\delta}(\boldsymbol{x}_1-\boldsymbol{x}_i)\bo{\delta}(\boldsymbol{p}_1-\boldsymbol{p}_i)\bo{\delta}(\boldsymbol{x}_2-\boldsymbol{x}_j)\bo{\delta}(\boldsymbol{p}_2-\boldsymbol{p}_j)\rangle$.  The expression (\ref{covar1}) above may be written in terms of these probability densities as:
\begin{eqnarray}
\langle\boldsymbol{u}(\boldsymbol{x})\bo{\cdot} \boldsymbol{u}(\boldsymbol{x}^\prime)\rangle & = & \int_{-L/2}^{L/2} f(s) \mathrm{d}s  \int_{-L/2}^{L/2}  f(s^\prime)\mathrm{d}s^\prime\nonumber\\ && \Bigg\{\int\mathrm{d}\boldsymbol{x}_1\mathrm{d}\boldsymbol{p}_1 [\mathsfbi{G}(\boldsymbol{x}-\boldsymbol{x}_1 - s\boldsymbol{p}_1)\bo{\cdot} \boldsymbol{p}_1]\bo{\cdot}[\mathsfbi{G}(\boldsymbol{x}^\prime-\boldsymbol{x}_1 - s\boldsymbol{p}_1)\bo{\cdot} \boldsymbol{p}_1]\Omega_1(\boldsymbol{x}_1, \boldsymbol{p}_1) \nonumber\\
&& \mbox{} \;\;+ \int\mathrm{d}\boldsymbol{x}_1\mathrm{d}\boldsymbol{p}_1\mathrm{d}\boldsymbol{x}_2\mathrm{d}\boldsymbol{p}_2[\mathsfbi{G}(\boldsymbol{x}-\boldsymbol{x}_1 - s\boldsymbol{p}_1)\bo{\cdot} \boldsymbol{p}_1]\bo{\cdot}[\mathsfbi{G}(\boldsymbol{x}^\prime-\boldsymbol{x}_2 - s\boldsymbol{p}_2)\bo{\cdot}\boldsymbol{p}_2]\nonumber\\
&&\mbox{}\;\;\;\qquad\Omega_2(\boldsymbol{r}, \boldsymbol{p}_1, \boldsymbol{p}_2)\Bigg\}.
\label{covar2}
\end{eqnarray}
Since we examine a spatially homogeneous isotropic suspension, one has $\Omega_1 = nL^3/(4\pi)$, and the pair probability density function $\Omega_2$ above only depends on the relative separation $\boldsymbol{r} = \boldsymbol{x}_2-\boldsymbol{x}_1$ of the pair of swimmers under consideration. The assumption of a statistically homogeneous and isotropic suspension of swimmers may only remain valid over a time smaller than \textit{O}$(nUL^2)^{-1}$. Evidence from simulations of spherical squirmers \citep{pagonabarraga2013, evans2011orientational, oyama2017hydrodynamically, alarcon2017morphology} does suggest that hydrodynamic interactions, even in a dilute setting, can induce global orientational order over times  long compared to the aforementioned scale. For slender swimmers, this assumption is not a limiting one, and the isotropic state appears to be stable to orientation perturbations at leading logarithmic order. We therefore proceed with the above expression for $\Omega_1$.

It is convenient to solve for the pair-probability density in Fourier space, and towards this end, the Fourier transformed covariance is expressible as:
\begin{eqnarray}
\langle\hat{\boldsymbol{u}}(\boldsymbol{k})\bo{\cdot} \hat{\boldsymbol{u}}(\boldsymbol{k}^\prime)\rangle & = & \bo{\delta}(\boldsymbol{k}+\boldsymbol{k}^\prime)\int_{-L/2}^{L/2} f(s) \mathrm{d}s  \int_{-L/2}^{L/2}  f(s^\prime)\mathrm{d}s^\prime\nonumber\\
&& \Bigg\{\frac{n}{4\upi}\int\mathrm{d}\boldsymbol{p}_1 [\hat{\mathsfbi{G}}(\boldsymbol{k})\bo{\cdot} \boldsymbol{p}_1]\bo{\cdot}[\hat{\mathsfbi{G}}(\boldsymbol{k})\cdot \boldsymbol{p}_1]\exp[-2\pi i \boldsymbol{k}\bo{\cdot}\boldsymbol{p}_1 (s-s^\prime)] \nonumber\\
&& \mbox{} \int\mathrm{d}\boldsymbol{p}_1\mathrm{d}\boldsymbol{p}_2[\hat{\mathsfbi{G}}(\boldsymbol{k})\bo{\cdot} \boldsymbol{p}_1]\bo{\cdot}[\hat{\mathsfbi{G}}(\boldsymbol{k})\cdot \boldsymbol{p}_2]\exp[-2\pi i (s\boldsymbol{k}\bo{\cdot}\boldsymbol{p}_1 - s^\prime\boldsymbol{k}\bo{\cdot}\boldsymbol{p}_2) ]\hat{\Omega}_2\Bigg\}.\nonumber\\
\label{covarFourier11}
\end{eqnarray}
Here, the Fourier transformed quantities are defined as $\hat{f}(\boldsymbol{k}) = \int \exp(-2\pi i \boldsymbol{k}\bo{\cdot}\boldsymbol{x}) f(\boldsymbol{x}) \mathrm{d}\boldsymbol{x}$, and the factor $\delta(\boldsymbol{k}+\boldsymbol{k}^\prime)$ represents the translational invariance due to the assumed spatial homogeneity. Using the aforementioned form of $f(s)$, and the Fourier transformed Oseen-Burgers tensor, $\hat{\mathsfbi{G}}(\boldsymbol{k}) = 1/(4\pi^2 \eta k^2) (\mathsfbi{I} - \hat{\bo{k}}\hat{\bo{k}})$, we get:
\begin{eqnarray}
\langle\hat{\boldsymbol{u}}(\boldsymbol{k})\bo{\cdot} \hat{\boldsymbol{u}}(\boldsymbol{k}^\prime)\rangle & = & \frac{\bo{\delta}(\boldsymbol{k}+\boldsymbol{k}^\prime)}{(\ln\kappa)^2}\Bigg\{\frac{nL^3}{16\pi^7 k^4}\int\mathrm{d}\boldsymbol{p}_1 \left[ \left( \mathsfbi{I} - \frac{\bo{k}\bo{k}}{k^2}\right)\right]\bo{:}\bo{p}_1\bo{p}_1\frac{\sin^4\left(\frac{\pi}{2}\boldsymbol{k}\bo{\cdot}\boldsymbol{p}_1\right)}{(\boldsymbol{k}\bo{\cdot}\boldsymbol{p}_1)^2} \nonumber\\
&& \mbox{} + \frac{1}{4\pi^6 k^2 k^{\prime 2}}  \int\mathrm{d}\boldsymbol{p}_1\mathrm{d}\boldsymbol{p}_2\left[ \left( \mathsfbi{I} - \frac{\bo{k}\bo{k}}{k^2}\right)\right]\bo{:}\bo{p}_1\bo{p}_2\nonumber\\
&& \mbox{}\qquad\qquad\qquad\;\; \frac{\sin^2\left(\frac{\pi}{2}\boldsymbol{k}\bo{\cdot}\boldsymbol{p}_1\right)}{\boldsymbol{k}\bo{\cdot}\boldsymbol{p}_1}    \frac{\sin^2\left(\frac{\pi}{2}\boldsymbol{k}\bo{\cdot}\boldsymbol{p}_2\right)}{\boldsymbol{k}\bo{\cdot}\boldsymbol{p}_2}\hat{\Omega}_2(\bo{p}_1, \bo{p}_2; \bo{k})\Bigg\},
\label{covarFourier3}
\end{eqnarray}
as the Fourier transformed velocity covariance in a suspension of slender swimmers. Note that (\ref{covarFourier3}) has been rendered non-dimensional by using $L$ and $U$, respectively, as length and velocity scales.

In what follows, we characterize the variance and covariance first for a suspension of non-interacting swimmers in \textsection\ref{subsec:noint}, and then, for suspensions of interacting straight and run-and-tumble swimmers, respectively, in \textsection\ref{subsubsec:intSt} and \textsection\ref{subsubsec:RTP}. For the latter case, we also present an alternate derivation of the covariance using a matched asymptotic expansions approach in \textsection\ref{subsubsec:matched_asymptotic}, applicable to RTP suspensions with large swimmer run lengths.

\subsection{Suspensions of non-interacting swimmers}\label{subsec:noint}
\begin{figure}
\begin{center}
\includegraphics[width=\textwidth]{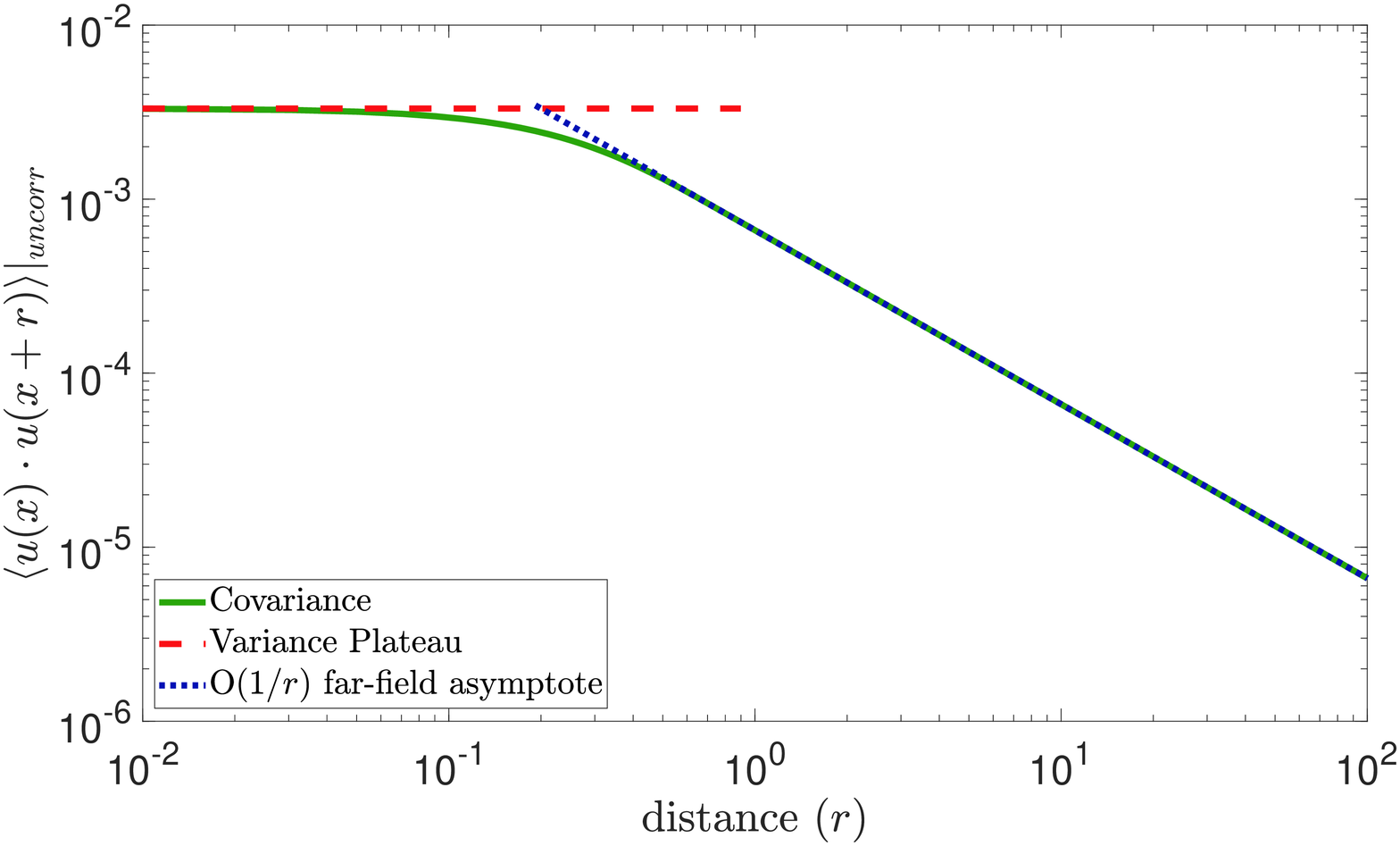}
\end{center}
\caption{The \textit{O}($nL^3$) velocity covariance in a dilute non-interacting suspension of slender swimmers. The covariance transitions from an \textit{O}($nL^3$)$/(\ln\kappa)^2$ variance plateau to a far-field \textit{O}($1/r$) asymptote.}
\label{fig2}
\end{figure}

In the absence of interactions, only the term involving the single swimmer probability density is relevant, so that (\ref{covarFourier3}) takes the simplified form:
\begin{equation}
\langle\hat{\bo{u}}(\bo{k})\cdot\hat{\bo{u}}(\bo{k}^\prime)\rangle = \frac{nL^3}{16\pi^7 (\ln\kappa)^2 k^4}\bo{\delta}(\bo{k}+\bo{k}^\prime)\int \mathrm{d}\bo{p}_1 \left[1 - \frac{\left(\bo{k}\bo{\cdot}\bo{p}_1\right)^2}{k^2}\right]\frac{\sin^4\left(\frac{\pi \bo{k}\bo{\cdot}\bo{p}_1}{2}\right)}{(\bo{k}\bo{\cdot}\bo{p}_1)^2}.
\label{covarSingle}
\end{equation}
An inverse Fourier transform of (\ref{covarSingle}) yields:
\begin{equation}
\langle\bo{u}(\bo{x})\cdot\bo{u}(\bo{x}+\bo{r})\rangle\vert_{uncorr} = \frac{nL^3}{4\pi^6 (\ln\kappa)^2 r}\int_0^\infty \mathrm{d}k \frac{\sin(2\pi kr)}{k^5}\int_{-1}^1\mathrm{d}\mu\left(\frac{1-\mu^2}{\mu^2}\right)\sin^4\left(\frac{\pi k \mu}{2}\right).
\label{covarSingleReal}
\end{equation}
Owing to isotropy, the covariance only depends on the scalar distance $r$. The variance is obtained on setting $r = 0$ in (\ref{covarSingleReal}), which gives $\langle\bo{u}(\bo{x})\cdot\bo{u}(\bo{x})\rangle\vert_{uncorr} = nL^3/[96\pi(\ln\kappa)^2]$. In the far-field ($r\gg 1$), we recover the limiting form of the covariance for point dipoles, given by: $\langle\bo{u}(\bo{x})\cdot\bo{u}(\bo{x}+\bo{r})\rangle\vert_{uncorr} = nL^3/[120\pi (\ln\kappa)^2 r]$ \citep{underhill2011}. The aforementioned variance and far-field covariance have been plotted alongside (\ref{covarSingleReal}), evaluated numerically, in figure \ref{fig2}. Note that the covariance, as given by (\ref{covarSingleReal}), depends only on the single-swimmer statistics at a given instant, and therefore, remains the same both for straight-swimmers and run-and-tumble swimmers provided the tumbles in the latter case are assumed to be instantaneous (so that any disturbance field generated during a tumble is neglected).

\subsection{Suspensions of interacting swimmers}\label{subsec:int}

Herein, we  consider the covariance in a suspension of hydrodynamically interacting slender swimmers to \textit{O}$(nL^3)^2$, with the contribution at this order requiring the analysis of pairwise interactions between slender swimmers. Therefore, the additional correlated contribution involving the pair-probability density in (\ref{covarFourier3}) needs to be calculated. This correlated contribution is given by:
\begin{eqnarray}
 \langle\hat{\bo{u}}(\bo{k})\bo{\cdot} \hat{\bo{u}}(\bo{k}^{\prime})\rangle\vert_{corr} &=  & \delta(\bo{k}+\bo{k}^\prime)\left\{ \frac{1}{4\pi^6(\ln\kappa)^2 k^2 k^{\prime 2}}  \int\mathrm{d}\bo{p}_1\mathrm{d}\bo{p}_2\left[ \left( \mathsfbi{I} - \frac{\bo{k}\bo{k}}{k^2}\right)\right]\bo{:}\bo{p}_1\bo{p}_2\nonumber\right.\\ 
 && \left.\frac{\sin^2\left(\frac{\pi}{2}\bo{k}\bo{\cdot}\bo{p}_1\right)}{\bo{k}\bo{\cdot}\bo{p}_1} \, \frac{\sin^2\left(\frac{\pi}{2}\bo{k}\bo{\cdot}\bo{p}_2\right)}{\bo{k}\bo{\cdot}\bo{p}_2}\hat{\Omega}_2\right\},
\label{eq:covarFourier5}
\end{eqnarray}
where $\hat{\Omega}_2$  is to be determined. 

Considering slender swimmers, with an aspect ratio $\kappa \gg1$, leads to logarithmically weak interactions on length scales of \textit{O}($L$) (scales that contribute dominantly to the velocity variance, as may be verified posteriori), and this allows one to expand $\Omega_2$ as a series in $(\ln\kappa)^{-1}$: $\Omega_2 = \Omega_2^{(0)} + 1/(\ln\kappa) \Omega_2^{(1)} + \ldots$; here, $\Omega_2^{(0)}$ is the product of two singlet probability densities, $(nL^3)^2/(4\pi)^2$, and denotes the uncorrelated base state at leading order. $\Omega_2^{(1)} \equiv \Omega_2^{(1)}(\bo{r}, \bo{p}_1, \bo{p}_2)$ gives the correlated probability, at \textit{O}($1/\ln\kappa$), of finding two swimmers with orientations $\bo{p}_1$ and $\bo{p}_2$, separated by $\bo{r}$. Typical bacteria such as \emph{E. coli} and \emph{B. subtilis} are quite slender \citep{Berg2004, clement2014}, and this assumption reasonably reproduces experimentally-measured disturbance velocity fields to distances of \textit{O}($L$) \citep{kasyapkoch2014}. Note that the contribution due to $\Omega_2^{(0)}$ in (\ref{eq:covarFourier5}) is identically zero. This is as expected since the velocity variance in an uncorrelated swimmer suspension must be proportional to $nL^3$, with no constraints of diluteness involved.

Therefore, to leading logarithmic order, one only needs to determine $\Omega_2^{(1)}(\bo{p}_1, \bo{p}_2; \bo{k})$ to characterize pair correlations in the swimmer suspension, and thence, determine the correlated contribution to the covariance in (\ref{eq:covarFourier5}), which is now  expressible as:
\begin{eqnarray}
   \langle\hat{\bo{u}}(\bo{k})\bo{\cdot}\hat{\bo{u}}(\bo{k}^\prime)\rangle\vert_{corr} =
   \frac{\bo{\delta}(\bo{k}+\bo{k}^\prime)}{4\pi^6(\ln\kappa)^3 k^4} && \left [ \int\mathrm{d}\bo{p}_1\mathrm{d}\bo{p}_2  \frac{1}{\bo{k}\bo{\cdot}\bo{p}_1}  \frac{1}{\bo{k}\bo{\cdot}\bo{p}_2} \sin^2(\frac{\pi}{2}\bo{k}\bo{\cdot}\bo{p}_1) \sin^2(\frac{\pi}{2}\bo{k}\bo{\cdot}\bo{p}_2)\right. \nonumber \\ 
   &&\left.  \;\left(\mathsfbi{I} - \frac{\bo{kk}}{k^2}\right)\bo{:}\bo{p}_1\bo{p}_2 \hat{\Omega}^{(1)}_2  \right ].
  \label{eq:fourier}
\end{eqnarray}
At this order, there is a crucial difference in the pair-correlations that develop in suspensions of straight-swimmers and RTPs, and therefore, the two cases are treated separately.


\subsubsection{Suspensions of interacting straight-swimmers}\label{subsubsec:intSt}

For straight-swimmers, $\Omega_N$ satisfies the Liouville equation, and evolves due to swimming, and due to convection and rotation of each swimmer by the disturbance velocity fields due to the remaining swimmers. In the dilute limit, integrating over the degrees of freedom of the remaining $N-2$ swimmers, while neglecting three-swimmer interactions, we obtain the equation for the pair probability density $\Omega_2$, at steady state, as:
\begin{equation}
  \bnabla_{\bo{r}}\bo{\cdot}\{[(U\bo{p}_2 + \bo{u}_{1}) - (U\bo{p}_1+\bo{u}_{2})]\Omega_2\}  \bnabla_{\bo{p}_1}\bo{\cdot}(\dot{\bo{p}}_{12} \Omega_2) + \bnabla_{\bo{p}_2}\bo{\cdot}(\dot{\bo{p}}_{21} \Omega_2) = 0.
  \label{eq:omegaST}
\end{equation}
The terms within braces in (\ref{eq:omegaST}) denote the convection of $\Omega_2$ by the relative velocity of the swimmer pair that includes contributions due to both swimming ($U\bo{p}_1$, $U\bo{p}_2$) and the disturbance velocity fields ($\bo{u}_{2}$, $\bo{u}_{1}$). The third and fourth terms denote rotation of the swimmer orientations due to the disturbance velocity fields, with $\dot{\bo{p}}_{ij}$ denoting the rotation of swimmer $i$ by the disturbance velocity field  due to swimmer $j$.


Using the aforementioned series expansion of $\Omega_2$ and neglecting the convection by the \textit{O}$(\ln \kappa)^{-1}$ disturbance velocity fields in (\ref{eq:omegaST}), at leading logarithmic order, pair-correlations develop along straight-swimming trajectories. Thus, at \textit{O}($1/\ln\kappa$), for hydrodynamically interacting slender swimmers, $\Omega_2^{(1)}$ satisfies:
\begin{equation}
(\bo{p}_2-\bo{p}_1)\bo{\cdot}\nabla_{\bo{r}}\Omega_2^{(1)} = - \frac{(nL^3)^2}{(4\pi)^2} \left[\nabla_{\bo{p}_{1}}\bo{\cdot}\dot{\bo{p}}_{12} + \nabla_{\bo{p}_2}\bo{\cdot}\dot{\bo{p}}_{21}\right ],
\label{omega21a}
\end{equation}
on applying the non-dimensionalization mentioned below (\ref{covarFourier3}). In Fourier space, (\ref{omega21a}) takes the form:
\begin{equation}
2\pi i\bo{k}\bo{\cdot}(\bo{p}_2-\bo{p}_1)\hat{\Omega}_2^{(1)} = - \frac{(nL^3)^2}{(4\pi)^2} \left[\nabla_{\bo{p}_1}\bo{\cdot}\hat{\dot{\bo{p}}}_{12} + \nabla_{\bo{p}_2}\bo{\cdot}\hat{\dot{\bo{p}}}_{21}\right].
\label{omega21b}
\end{equation}
The solution for $\hat{\Omega}_2^{(1)}$ from (\ref{omega21b}) is expressible as:
\begin{equation}
\hat{\Omega}_{2}^{(1)} = c\,\delta\left[\bo{k}\bo{\cdot}(\bo{p}_2-\bo{p}_1)\right] - \frac{(nL^3)^2}{(4\pi)^2 2\pi i \bo{k}\bo{\cdot}(\bo{p}_2-\bo{p}_1)}\left[\nabla_{\bo{p}_1}\bo{\cdot}\hat{\dot{\bo{p}}}_{12} + \nabla_{\bo{p}_2}\bo{\cdot}\hat{\dot{\bo{p}}}_{21}\right],
\label{omega21c}
\end{equation}
where the first term on the right side, involving the Dirac delta function, is the homogeneous solution, and the expressions for $\hat{\dot{\bo{p}}}_{12}$ and $\hat{\dot{\bo{p}}}_{21}$ that appear in the particular solution are given in appendix \ref{sec:appRotationRate}. The constant $c$ is determined from the fact that the swimmers are uncorrelated prior to interaction; that is, $\Omega_2^{(1)} \to 0$ for swimmers when they are infinitely separated in the upstream direction. Choosing a Cartesian coordinate system with the $z$ axis along the relative swimming velocity vector, that is $\hat{\bo{z}} = (\bo{p}_2-\bo{p}_1)/|\bo{p}_2-\bo{p}_1)|$, implies that $z \to-\infty$ denotes upstream infinity. Therefore, the aforementioned constraint of uncorrelated swimmers infinitely far upstream may be given as:
\begin{equation}
\lim_{z\to-\infty}\Omega_2^{(1)}  \equiv\lim_{z\to-\infty}\int \mathrm{d}\bo{k}\exp\left[2\pi i \bo{k}\bo{\cdot}\bo{x}\right]\hat{\Omega}_2^{(1)} = 0.
\label{omega21d_1}
\end{equation} 
One can represent $\bo{x} = (\bo{x}^\perp, z)\equiv (\bo{x}\bo{\cdot}[\mathsfbi{I}-\hat{\bo{z}}\hat{\bo{z}}], z)$, and since (\ref{omega21d_1}) is valid for arbitrary $\bo{k}^\perp$, one need only consider the integral over $k_z$, whence one obtains:
\begin{eqnarray}
\lim_{z\to-\infty}&\int& \mathrm{d}k_z\exp\left[2\pi i k_z z\right]\nonumber\\
&&\left[c \frac{\delta[k_z |\bo{p}_2-\bo{p}_1|]}{|\bo{p}_2-\bo{p}_1|} - \frac{(nL^3)^2}{(4\pi)^2 2\pi i k_z|\bo{p}_2-\bo{p}_1|}\left[\nabla_{\bo{p}_1}\bo{\cdot}\hat{\dot{\bo{p}}}_{12} + \nabla_{\bo{p}_2}\bo{\cdot}\hat{\dot{\bo{p}}}_{21} \right]\right] = 0.\nonumber\\
\label{omega21d}
\end{eqnarray}
While the first term within brackets in (\ref{omega21d}) may be evaluated readily, for $z\to-\infty$, the dominant contribution to the second term arises when $k_z\rightarrow 0$ such that $k_z z$ remains finite. This results in the following expression for $c$:
\begin{equation}
c = \frac{(nL^3)^2}{32\pi^2 |\bo{p}_2-\bo{p}_1|} \lim_{k_z\to 0}\left[\nabla_{\bo{p}_1}\bo{\cdot}\hat{\dot{\bo{p}}}_{12} + \nabla_{\bo{p}_2}\bo{\cdot}\hat{\dot{\bo{p}}}_{21} \right].
\label{omega21e}
\end{equation}
We now choose $\bo{x}^\perp = (x, y)$ such that $\hat{\bo{x}} = (\bo{p}_2\wedge\bo{p}_1)/|\bo{p}_2\wedge\bo{p}_1|$ and $\hat{\bo{y}} =(\bo{p}_2+\bo{p}_1)/|\bo{p}_2+\bo{p}_1|$, are the unit vectors in the plane orthogonal to $\hat{\bo{z}}$. For $k_z\to 0$, we have, $\bo{k}\bo{\cdot} \bo{p}_{1} = \bo{k}\bo{\cdot} \bo{p}_{2}$ with each of them being given by $k_y|\bo{p}_2+\bo{p}_1|/2$. Using these relations in (\ref{omega21e}), and after some algebra, $c$ is given by:

\begin{eqnarray}
c &=& \frac{3 (nL^3)^2}{8 \pi^5}\left[\frac{|\bo{p}_2-\bo{p}_1|^2}{[\bo{k}\wedge(\bo{p}_2-\bo{p}_1)]\bo{\cdot}[\bo{k}\wedge(\bo{p}_2-\bo{p}_1)]}\right] \frac{1}{\left(\bo{k}\bo{\cdot}(\bo{p}_1+\bo{p}_2)\right)} \sin^2\left(\frac{\pi}{4}\bo{k}\bo{\cdot}(\bo{p}_1+\bo{p}_2)\right)\nonumber\\
&& \sin\left(\frac{\pi}{2}\bo{k}\bo{\cdot}(\bo{p}_1+\bo{p}_2)\right)\left[\bo{p}_2\bo{\cdot}\bo{p}_1 - \frac{\left(\bo{k}\bo{\cdot}(\bo{p}_1+\bo{p}_2)\right)^2|\bo{p}_2-\bo{p}_1|^2}{4\left(\bo{k}\,\wedge(\bo{p}_2-\bo{p}_1)\right)\bo{\cdot}\left(\bo{k}\,\wedge(\bo{p}_2-\bo{p}_1)\right)}\right].
\label{c_final}
\end{eqnarray}
From (\ref{pdot}) and (\ref{c_final}) we finally get the \textit{O}($1/\ln \kappa$) correction to the Fourier transformed pair-probability density, in (\ref{omega21c}), as:
\begin{eqnarray}
   \hat{\Omega}_2^{(1)} &=& \frac{3 (nL^3)^2}{8\pi^5}\left[\frac{|\bo{p}_2-\bo{p}_1|^2}{\left(\bo{k}\wedge(\bo{p}_2-\bo{p}_1)\right)\bo{\cdot}\left(\bo{k}\wedge(\bo{p}_2-\bo{p}_1)\right)}\right] \frac{1}{\left(\bo{k}\bo{\cdot}(\bo{p}_1+\bo{p}_2)\right)} \sin^2\left(\frac{\pi}{4}\bo{k}\bo{\cdot}(\bo{p}_1+\bo{p}_2)\right)  \nonumber \\ 
&& \sin\left(\frac{\pi}{2}\bo{k}\bo{\cdot}(\bo{p}_1+\bo{p}_2)\right)\!\!\!\left[\bo{p}_2\bo{\cdot}\bo{p}_1 - \frac{\left(\bo{k}\bo{\cdot}(\bo{p}_1+\bo{p}_2)\right)^2|\bo{p}_2-\bo{p}_1|^2}{4\left(\bo{k}\wedge(\bo{p}_2-\bo{p}_1)\right)\bo{\cdot}\left(\bo{k}\wedge(\bo{p}_2-\bo{p}_1)\right)}\right] \bo{\delta}\left[\bo{k}\bo{\cdot}(\bo{p}_2-\bo{p}_1)\right]\nonumber\\
&& - \frac{3 i (nL^3)^2}{32\pi^6 k^3}PV\!\!\left[\frac{1}{\hat{\bo{k}}\bo{\cdot}(\bo{p}_2-\bo{p}_1)}\right] \!\!\left(\bo{I}-\hat{\bo{k}}\hat{\bo{k}}\right)\!\!\bo{:}\bo{p}_1\bo{p}_2\!\! \left[\frac{1}{\left(\bo{k}\bo{\cdot}\bo{p}_1\right)} \sin^2\left(\frac{\pi}{2}\bo{k}\bo{\cdot}\bo{p}_1\right)\sin\left(\pi \bo{k}\bo{\cdot}\bo{p}_2\right)\right. \nonumber\\
&&\left.\qquad\qquad\qquad\qquad\qquad\qquad\qquad\qquad\qquad \frac{1}{\left(\bo{k}\bo{\cdot}\bo{p}_2\right)} \sin^2\left(\frac{\pi}{2}\bo{k}\bo{\cdot}\bo{p}_2\right)\sin\left(\pi \bo{k}\bo{\cdot}\bo{p}_1\right) \right] ,\nonumber\\
  \label{eq:omega21Full}
\end{eqnarray}
where the first term on the right side of (\ref{eq:omega21Full}) is the homogeneous solution, with $c$ being replaced by (\ref{c_final}); the integral involving the argument of $PV[\cdot]$ needs to be interpreted as a Cauchy principal value integral. The analysis of the homogeneous solution above is crucial, since it is this term alone that contributes to the covariance.


On using (\ref{eq:omega21Full}) in (\ref{eq:fourier}), and on applying the inverse Fourier transform, we obtain the \textit{O}$(nL^3)^2$ fluid velocity covariance in an interacting straight-swimmer suspension to be:
\begin{eqnarray}
\langle\bo{u}(\bo{x})\bo{\cdot}\bo{u}(\bo{x}+\bo{r})\rangle\vert_{corr} = \frac{3(nL^3)^2}{128\pi^6(\ln\kappa)^3 r}\int_0^\infty \mathrm{d}k \frac{\sin(2\pi kr)}{k^6} &&\int_{-1}^1 \mathrm{d}\mu (1-\mu^2)^2 \sin(\pi k \mu)\nonumber\\
&&  \frac{1}{\left(\frac{\pi}{2}k\mu\right)^3}\sin^6\left(\frac{\pi}{2}k\mu\right).
\label{eq:stcovar}
\end{eqnarray}
Again, the variance is obtained by setting $r=0$ in (\ref{eq:stcovar}), which gives:
\begin{equation}
\langle\bo{u}(\bo{x})\bo{\cdot}\bo{u}(\bo{x})\rangle\vert_{corr} = \frac{3(nL^3)^2}{64\pi^5(\ln\kappa)^3}\int_0^\infty \mathrm{d}k \frac{1}{k^5} \int_{-1}^1 \mathrm{d}\mu (1-\mu^2)^2 \sin(\pi k \mu) \frac{1}{\left(\frac{\pi}{2}k\mu\right)^3}\sin^6\left(\frac{\pi}{2}k\mu\right).
\label{eq:stvar}
\end{equation}
Notice that the integrand for $k$ in (\ref{eq:stvar}) scales as $1/k$ as $k\to 0$, implying a logarithmic  divergence. Thus, the \textit{O}$(nL^3)^2$ contribution to the covariance is not well defined for any $r$, and in particular, results in a logarithmically divergent variance on setting $r=0$. The physical arguments for this logarithmic divergence may now be laid out in a little more detail than in the introduction. As mentioned therein, the divergence arises from the slow\textit{O}($1/r^2$) decay of the pair-probability density in the far-field. This can be readily inferred from (\ref{eq:omega21Full}), where power counting in $k$, as $k\to 0$ suggests that $\hat{\Omega}_2^{(1)}\sim$ \textit{O}($1/k$) (which in real space would imply $\Omega_2^{(1)}\sim$ \textit{O}($1/r^2$) in the far-field). Alternatively, one may also observe this from the physical space form, given by (\ref{omega21a}), wherein integrating along the trajectory of relative swimming, that is: $\Omega_{2}^{(1)} = -(nL^3)^2/[(4\pi)^2U|\bo{p}_2-\bo{p}_1|]\int_{-\infty}^z \mathrm{d}z^\prime (\bnabla_{\bo{p}_1}\bo{\cdot}\dot{\bo{p}}_{12} + \bnabla_{\bo{p}_2}\bo{\cdot}\dot{\bo{p}}_{21})$, with $\dot{\bo{p}}_{ij}\sim$ \textit{O}($1/r^3$), the velocity gradient scaling in the far-field, leads to the \textit{O}($1/r^2$) decay. In this scenario, where a given swimmer interacts pair-wise with a cloud of swimmers surrounding it (see figure \ref{fig:schematic}b), one has the fluid velocity variance due to a \enquote{correlated cloud}, of radius $r$, scaling as $[\bo{u}(\bo{x})\bo{\cdot}\bo{u}(\bo{x})]_{cloud} \equiv nL^3\int U^2 r^{\prime 2}\mathrm{d}r^\prime/(r^4 r^{\prime 2}) \sim nL^3 U^2/r^3$, which when integrated over all such correlation clouds in position space, gives $\langle\bo{u}(\bo{x})\bo{\cdot}\bo{u}(\bo{x})\rangle|_{corr} \equiv (nL^3)^2 \ln(L_{box}/L)$; a logarithmically divergent variance.

\begin{figure}
\begin{center}
\includegraphics[width=\textwidth]{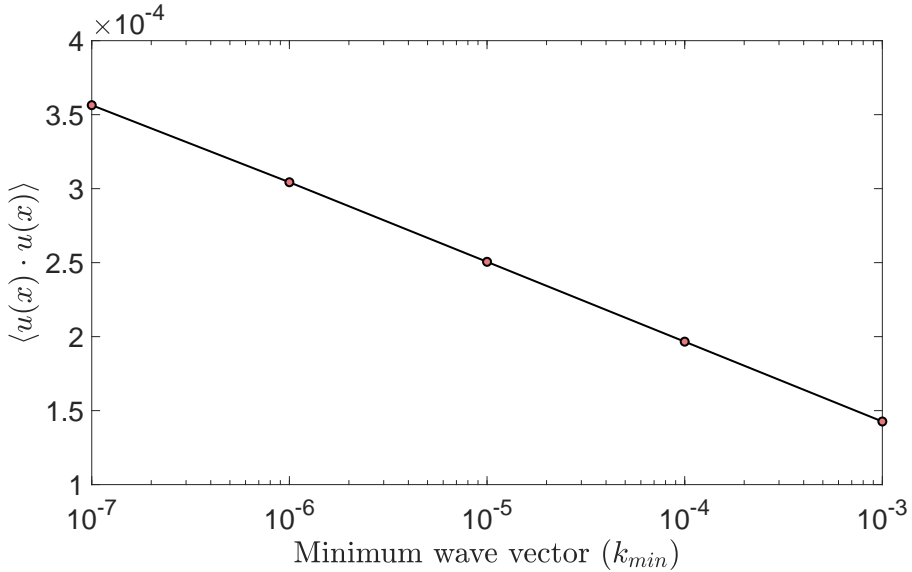}
\end{center}
\caption{The \textit{O}$(nL^3)^2$ correlated straight-swimmer variance plotted against the infrared cut-off ($k_{min}$) used to evaluate (\ref{eq:stvar}). Expectedly, it exhibits a logarithmic divergence with decreasing $k_{min}$.}
\label{fig3}
\end{figure}

In practice, the logarithmic divergence above  will be cut off at an appropriate screening length $L_{screen}$, such that the total variance, including the uncorrelated, given by (\ref{covarSingleReal}), and the correlated contribution above, takes the form $c_1 (nL^3) + (nL^3)^2\{c_2 \ln[L_{screen}/L] + c_3\}$, with the constants $c_i$'s being functions of swimmer aspect ratio. The screening length depends on the particular scenario. In box-size-limited simulations, the largest admissible wavelength is set by the computational domain, so $L_{screen} = k_{min}^{-1}\equiv L_{box}$. In figure \ref{fig3}, we highlight this box-size-dependent divergence of the correlated contribution as a function of $k_{min}$, with $k_{min}^{-1}$ replacing 0 as the lower limit of the Fourier integral in (\ref{eq:stvar}), and thereby, enforcing a long wavelength cut-off. For suspensions of interacting straight-swimmer, the box-size limitation arises provided the swimmer mean free path, which is \textit{O}$(nL^2)^{-1}$, is larger than $L_{box}$. For $L_{box}\gg(nL^2)^{-1}$, these suspensions become linearly unstable to orientation perturbations, and one expects a transition to collective motion \citep{saintillan2008, subkoch2009}; in simulations, this is characterized by a box size dependent sharp increase in the fluid velocity variance and tracer diffusivities \citep{saintillan2011emergence, deepak2015, stenhammar2017}. As inherent in our homogeneous suspension assumption, our analysis therefore caters to the former scenario (upto $L_{box}\sim(nL^2)^{-1}$ or for time scales $\sim$ \textit{O}$(nUL^2)^{-1}$ for larger box sizes). Simulations of interacting regularized force dipole swimmers carried out by  \cite{underhill2011}, at $nL^3\sim10^{-2}$ (and in the regime $L_{box}<(nL^2)^{-1}$), do point to a logarithmic divergence of the variance, as may be inferred from the \textit{O}($1/r^2$) far-field decay of the spatial orientation correlations, and thence, the pair-probability density (see figure 7 therein). Interestingly, the authors observe  a logarithmically diverging variance in simulations, even at higher volume fractions ($nL^3\sim 10^{-1}$),  for a range of box sizes $L_{box}\geq (nL^2)^{-1}$. For slender swimmers, owing to the weak pair correlations (at least by a factor of $\ln\kappa$), the orientation decorrelation is expected to occur over a logarithmically large number of pair interactions, implying a larger effective mean free path. Thus, our analysis may remain valid for suspension of slender swimmers at even higher volume fractions than for the regularized point dipoles considered in earlier simulations \citep{G2008, underhill2011, stenhammar2017}.

For real bacteria, intrinsic decorrelation mechanisms such as rotary diffusion or run-and-tumble dynamics, lead to $L_{screen}$ = $U/D_r$ or $U\tau$, where $D_r$ is the rotary diffusivity and $\tau$ the mean run duration. Note that the rotary diffusion above could have an entirely hydrodynamic origin. For slender straight-swimmer suspensions not limited by box size, one expects the logarithmic divergence to nevertheless be cut off at $L_{screen} \sim$ \textit{O}$(U/D_r^h)$, where $D_r^h \sim$ \textit{O}$(nUL^2/(\ln\kappa)^2)$ is a hydrodynamic rotary diffusivity arising from logarithmically weak pairwise interactions between slender swimmers each of which lead to an \textit{O}($1/\ln \kappa$) angular displacement. This diffusivity has been calculated earlier \citep{subkoch2009}, and accounting for this finite $D_r^h$ leads to a suspension velocity variance of the form: $c_1 (nL^3) + (nL^3)^2\{c_2 \ln[(\ln\kappa)^2/(n L^3)] + c_3\}$. As will be seen in the next subsection, the covariance integral, (\ref{eq:stcovar}), in the strict straight-swimmer limit ($U\tau/L = \infty$) is non-decaying. However, the straight-swimmer limit is better interpreted as a limiting case of RTPs, in which case one finds an intermediate asymptotic regime where the covariance exhibits a weak logarithmic decay with separation.

\subsubsection{Suspensions of interacting RTPs}\label{subsubsec:RTP}
The kinetic equation for the pair-probability density for a suspension of RTPs includes additional terms compared to (\ref{omega21a}) that describe the tumble dynamics of the individual bacteria, and is given by:
\begin{eqnarray}
\left(\frac{U\tau}{L}\right)(\bo{p}_2-\bo{p}_1)\bo{\cdot}\bnabla_{\bo{r}}\Omega_2^{(1)} &+& \left(\Omega_2^{(1)} -\frac{1}{4\pi}\int\mathrm{d}\bo{p}_1 \Omega_2^{(1)}\right) + \left(\Omega_2^{(1)} -\frac{1}{4\pi}\int\mathrm{d}\bo{p}_2 \Omega_2^{(1)}\right)\nonumber\\
&&  = - \frac{(nL^3)^2}{(4\pi)^2} \left(\frac{U\tau}{L}\right) \left[\bnabla_{\bo{p}_1}\bo{\cdot}\dot{\bo{p}}_{12} + \bnabla_{\bo{p}_2}\bo{\cdot}\dot{\bo{p}}_{21}\right ].
\label{omega21RTP1}
\end{eqnarray}
The bracketed terms on the left-hand side of (\ref{omega21RTP1}) correspond to the swimmers undergoing tumbles in accordance with Poisson statistics \citep{subkoch2009, othmer1988models}, with there being no correlation between pre- and post-tumble orientations (random tumbles); $\tau$ is the mean run duration (that is, the mean interval between successive random tumbles). In the absence of additional forcing, tumbling causes the single-swimmer distribution to relax to isotropy on a time scale of \textit{O}($\tau$). Here, $U\tau/L$ is a non-dimensional mean run length; thus, $U\tau/L \rightarrow \infty$ in the limit of straight-swimmers. Again, on Fourier transforming, one gets:
\begin{eqnarray}
\left(\frac{U\tau}{L}\right)2\pi i \bo{k}\bo{\cdot}(\bo{p}_2-\bo{p}_1)\hat{\Omega}_2^{(1)} &+& \left(2 \hat{\Omega}_2^{(1)} -\frac{1}{4\pi}\int\mathrm{d}\bo{p}_1 \hat{\Omega}_2^{(1)} -\frac{1}{4\pi}\int\mathrm{d}\bo{p}_2 \hat{\Omega}_2^{(1)}\right) \nonumber \\
&&= - \frac{(nL^3)^2}{(4\pi)^2}\left(\frac{U\tau}{L}\right) \left[\bnabla_{\bo{p}_1}\bo{\cdot}\hat{\dot{\bo{p}}}_{12} + \bnabla_{\bo{p}_2}\bo{\cdot}\hat{\dot{\bo{p}}}_{21}\right].
\label{omega21RTP2}
\end{eqnarray}
One can obtain an analytical solution for $\hat{\Omega}_2^{(1)}(\bo{p}_1, \bo{p}_2; \bo{k})$ in (\ref{omega21RTP2}), by determining the Green's function of the linear operator on the left-hand side. This requires one to first solve the time dependent operator, and then obtain the steady state pair probability density by taking the long time limit. The solution procedure involves representing the Green's function as a superposition of the eigenfunctions. The singular nature of the linear operator implies that both the discrete and (singular) continuous spectrum need to be examined. For the sake of brevity, we will be reporting this detailed Green's function analysis elsewhere  \citep{ganesh20}.

For purposes of the covariance calculation, it is worth noting that the symmetry of the orientation dilatation forcing terms on the right-hand side of (\ref{omega21RTP2}) ensures that one does not require information of the full  Green's function. A representation of $\hat{\Omega}_2^{(1)}$ as a convolution of the Green's function with the right-hand side of (\ref{omega21RTP2}) (as briefly mentioned above) implies that only those eigenfunctions, that the orientation dilatation terms project onto, contribute. To see this, we first choose a coordinate system with its polar axis along $\hat{\bo{k}}$, such that $\mu_i \equiv \cos\theta_i = \hat{\bo{k}}\bo{\cdot}\bo{p}_i$, with $\theta_i$ being the polar angle, and $\phi_i$ representing the azimuthal angle measured in a plane perpendicular to $\bo{k}$. In this coordinate system, the orientation dilatation terms are proportional to $\cos\phi_i$, and hence, vanish when integrated in orientation space. Thus, one need only solve for a reduced pair probability density, which accounts for eigen modes proportional to $\cos\phi_i$. Now, eigen modes proportional to $\cos(n \phi_i)$ or $\sin(m\phi_i)$ remain unaffected by the integral terms for $m,n\neq 0$, since the latter vanish for all such cases; physically, the orientation dilatation terms lead to pure-orientation correlations without concentration (number density) perturbations. Thus, one may set $\int\hat{\Omega}_2^{(1)}\mathrm{d}\bo{p}_1 = \int\hat{\Omega}_2^{(1)}\mathrm{d}\bo{p}_2 = 0$, and the governing equation for $\hat{\Omega}_2^{(1)}$ reduces to:
\begin{equation}
\left(\frac{U\tau}{L}\right)2\pi i \bo{k}\bo{\cdot}(\bo{p}_2-\bo{p}_1)\hat{\Omega}_2^{(1)} + 2 \hat{\Omega}_2^{(1)} = - \frac{(nL^3)^2}{(4\pi)^2}\left(\frac{U\tau}{L}\right) \left[\bnabla_{\bo{p}_1}\bo{\cdot}\hat{\dot{\bo{p}}}_{12} + \bnabla_{\bo{p}_2}\bo{\cdot}\hat{\dot{\bo{p}}}_{21}\right].
\label{omega21RTPModified}
\end{equation}
Solving (\ref{omega21RTPModified}), yields the following exact expression:
\begin{eqnarray}
   \hat{\Omega}_2^{(1)} &=&  \frac{3 (nL^3)^2}{32\pi^5 k^2}\left(\frac{U\tau}{L}\right)\left(\frac{1}{\pi i (U\tau/L)\bo{k}\bo{\cdot}(\bo{p}_2-\bo{p}_1) + 1}\right) \left(\mathsfbi{I}-\hat{\bo{k}}\hat{\bo{k}}\right)\bo{:}\bo{p}_2\bo{p}_1 \nonumber\\
 &&\left[\frac{1}{\left(\bo{k}\bo{\cdot}\bo{p}_1\right)} \sin^2\left(\frac{\pi}{2}\bo{k}\bo{\cdot}\bo{p}_1\right)\sin\left(\pi \bo{k}\bo{\cdot}\bo{p}_2\right) +  \frac{1}{\left(\bo{k}\bo{\cdot}\bo{p}_2\right)} \sin^2\left(\frac{\pi}{2}\bo{k}\bo{\cdot}\bo{p}_2\right)\sin\left(\pi \bo{k}\bo{\cdot}\bo{p}_1\right) \right] .\nonumber\\
  \label{eq:omega21RTP5}
\end{eqnarray}
The pair-correlation function for RTPs above does not involve any concentration fluctuations, as may seen by integrating (\ref{eq:omega21RTP5}) over orientation space. From (\ref{eq:omega21RTP5}), one can also infer that for large $U\tau/L$ (the straight-swimmer limit), $\hat{\Omega}_2^{(1)}\sim$ \textit{O}($1/k$) as $k\to 0$, implying that $\Omega_2^{(1)} \sim$ \textit{O}($1/r^2$) in the far-field; however, for rapid tumblers ($U\tau/L\ll 1$), $\Omega_2^{(1)}$ exhibits a more rapid far-field decay than an algebraic one. The polar and nematic correlations between swimmer orientations may be derived using the above expression for $\hat{\Omega}_2^{(1)}$ and are discussed in Appendix \ref{sec:orientation}.

One may recover the straight-swimmer limiting case, obtained in \textsection\ref{subsubsec:intSt}, from (\ref{eq:omega21RTP5}), in the limit of large $U\tau/L$.  To see this, we first rewrite (\ref{eq:omega21RTP5}) as:
\begin{eqnarray}
   \hat{\Omega}_2^{(1)} &=& - \frac{3 (nL^3)^2}{32\pi^6 k^3}\left(\frac{i}{\hat{\bo{k}}\bo{\cdot}(\bo{p}_2-\bo{p}_1) + i\vartheta}\right) \left(\mathsfbi{I}-\hat{\bo{k}}\hat{\bo{k}}\right)\bo{:}\bo{p}_2\bo{p}_1 \nonumber\\
 &&\left[\frac{1}{\left(\bo{k}\bo{\cdot}\bo{p}_1\right)} \sin^2\left(\frac{\pi}{2}\bo{k}\bo{\cdot}\bo{p}_1\right)\sin\left(\pi \bo{k}\bo{\cdot}\bo{p}_2\right) +  \frac{1}{\left(\bo{k}\bo{\cdot}\bo{p}_2\right)} \sin^2\left(\frac{\pi}{2}\bo{k}\bo{\cdot}\bo{p}_2\right)\sin\left(\pi \bo{k}\bo{\cdot}\bo{p}_1\right) \right] ,\nonumber\\
  \label{eq:omega21RTP5Recovery1}
\end{eqnarray}
where $\vartheta = L/(\pi k U\tau)$. In the limit $U\tau/L\gg 1$, $\vartheta\ll 1$, implying that $i\vartheta\to 0$. Interpreting the bracketed term on the right-hand side of (\ref{eq:omega21RTP5Recovery1}), in this limit, needs careful consideration. Naively setting $\vartheta = 0$ only leads to the particular solution corresponding to the straight-swimmer pair probability density, as given in (\ref{eq:omega21Full}). To obtain the correct answer in the straight-swimmer limiting form of $\hat{\Omega}_2^{(1)}$, one needs to apply the Plemelj-Sokhotski formula \citep{Ghakov1966, AblowitzFokas}, which may written as $\lim_{\epsilon \to 0} 1/(x+ i\epsilon) = -\delta(x) + PV[1/x]$, and this yields

\begin{eqnarray}
   \hat{\Omega}_2^{(1)}\vert_{st} &=& - \frac{3 (nL^3)^2}{32\pi^6 k^3}\left(\bo{\delta}(\hat{\bo{k}}\bo{\cdot}(\bo{p}_2-\bo{p}_1))-PV\left[\frac{i}{\hat{\bo{k}}\bo{\cdot}(\bo{p}_2-\bo{p}_1)}\right]\right) \left(\mathsfbi{I}-\hat{\bo{k}}\hat{\bo{k}}\right)\bo{:}\bo{p}_2\bo{p}_1 \nonumber\\
 &&\left[\frac{1}{\left(\bo{k}\bo{\cdot}\bo{p}_1\right)} \sin^2\left(\frac{\pi}{2}\bo{k}\bo{\cdot}\bo{p}_1\right)\sin\left(\pi \bo{k}\bo{\cdot}\bo{p}_2\right) +  \frac{1}{\left(\bo{k}\bo{\cdot}\bo{p}_2\right)} \sin^2\left(\frac{\pi}{2}\bo{k}\bo{\cdot}\bo{p}_2\right)\sin\left(\pi \bo{k}\bo{\cdot}\bo{p}_1\right) \right] .\nonumber\\
  \label{eq:omega21RTP5Recovery2}
\end{eqnarray}
One may now readily note that the term involving $\bo{\delta}(\hat{\bo{k}}\bo{\cdot}(\bo{p}_2-\bo{p}_1))$ in (\ref{eq:omega21RTP5Recovery2}) constitutes the homogeneous solution, whereas, the term involving $PV[\cdot]$ is the particular solution of the straight-swimmer pair probability density given by (\ref{eq:omega21Full}).

\begin{figure}
\includegraphics[width=\textwidth]{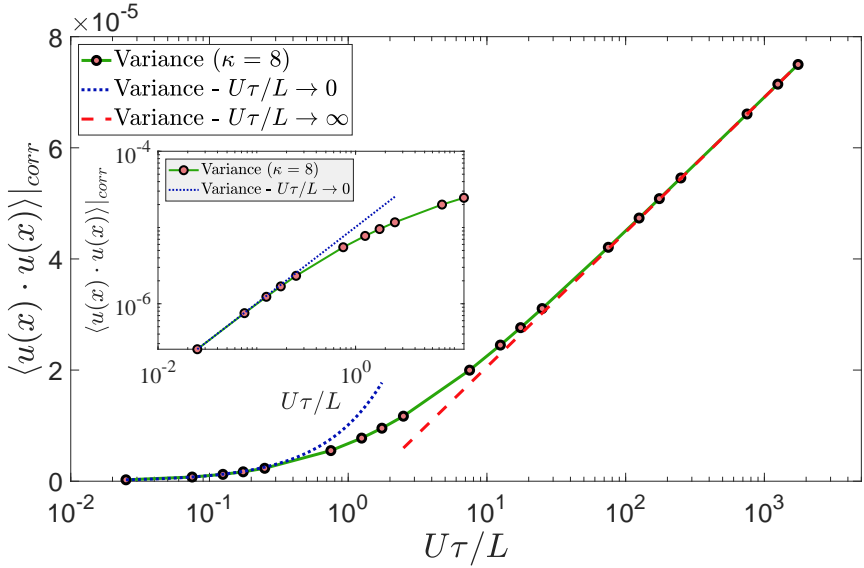}
\caption{\label{fig4} The \textit{O}$(nL^3)^2$ correlated variance, as given by (\ref{varRTP_corr}), plotted as a function of $U\tau/L$. The small (dotted) and large (dash-dotted) $U\tau/L$ asymptotes are shown; the inset highlights the linear scaling of the correlated variance with $U\tau/L$ for $U\tau/L\ll 1$.}
\end{figure}

\begin{figure}
\includegraphics[width=\textwidth]{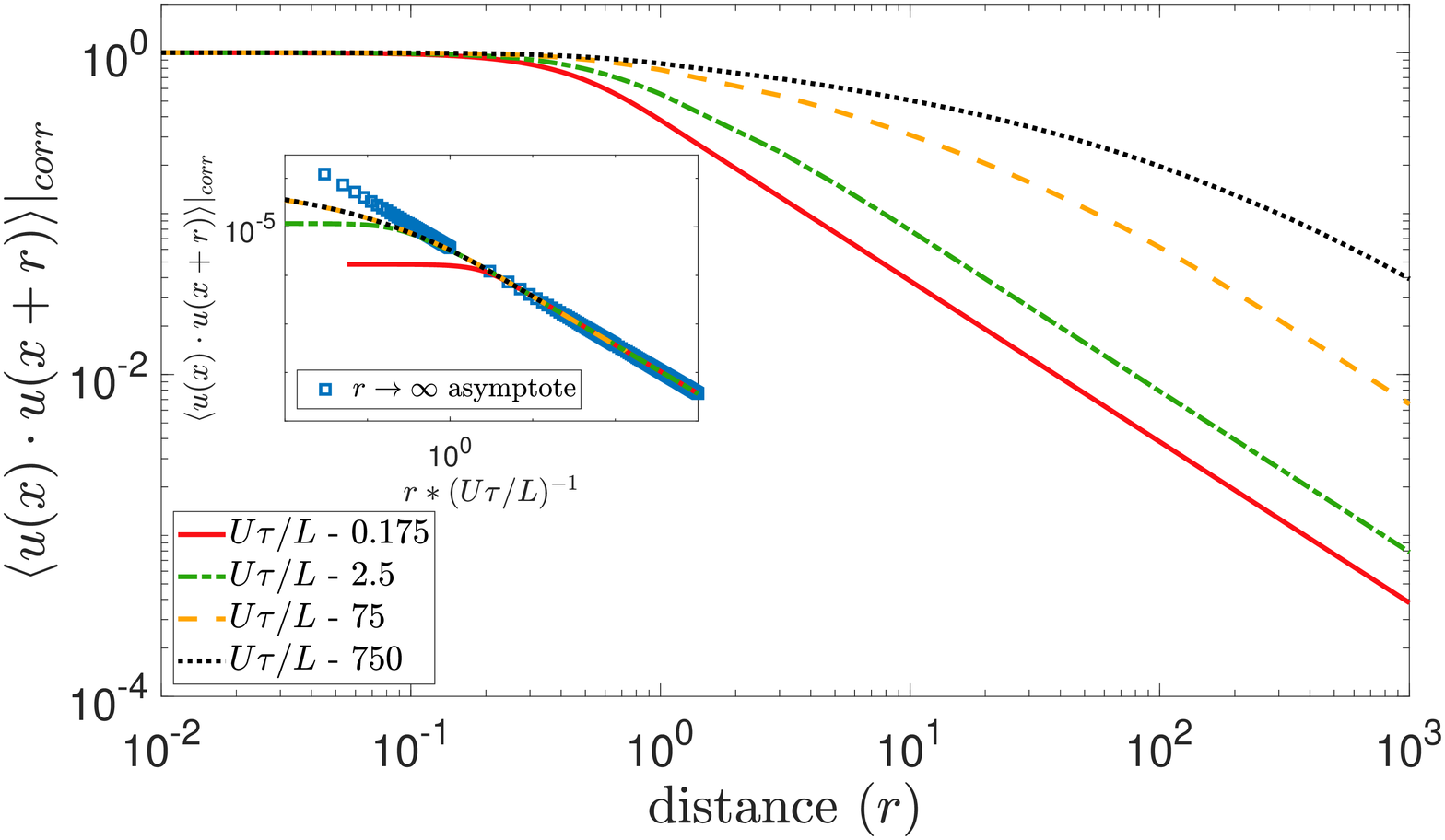}
\caption{\label{fig5} The \textit{O}$(nL^3)^2$ component of the (normalized) fluid velocity covariance as a function of $r$. In the inset the covariance curves collapse onto the far-field dipole asymptote  (represented by the $\square$ symbols) stated below (\ref{covarRTP}), for $r\gg U\tau/L$, when the abscissa is scaled with $(U\tau/L)^{-1}$ .}
\end{figure}

Now, using the coordinate system described above, the Fourier transformed covariance from (\ref{eq:fourier}) is expressible as:
\begin{eqnarray}
   \langle\hat{\bo{u}}(\bo{k})\bo{\cdot}\hat{\bo{u}}(\bo{k}^\prime)\rangle\vert_{corr} &=&  
 \frac{(nL^3)^2}{4\pi^6(\ln\kappa)^3} \int_0^{2\pi}\mathrm{d}\phi_1\int_0^{2\pi}\mathrm{d}\phi_2 \int_{-1}^1\mathrm{d}\mu_1 \nonumber\\ 
 &&\int_{-1}^1\mathrm{d}\mu_2 \frac{1}{k\mu_1}\sin^2\left(\frac{\pi}{2}k\mu_1\right)\frac{1}{k\mu_2}\sin^2\left(\frac{\pi}{2}k\mu_2\right)\nonumber\\
 &&\qquad\left(1-\mu_1^2\right)^{1/2}\left(1-\mu_2^2\right)^{1/2}\cos(\phi_2-\phi_1)\hat{\Omega}_2^{(1)}.
  \label{eq:fourierRTP}
\end{eqnarray}
On inverse Fourier transforming (\ref{eq:fourierRTP}) one obtains the correlated contribution to the fluid velocity covariance for interacting RTPs to be:
\begin{eqnarray}
\langle\bo{u}(\bo{x})\bo{\cdot}\bo{u}(\bo{x} + \bo{r})\rangle\vert_{corr} &=&  \frac{3(nL^3)^2}{128\pi^6(\ln\kappa)^3 r}\left(\frac{U\tau}{L}\right) \int_0^\infty \mathrm{d}k \frac{\sin(2\pi k r)}{k^5}\int_{-1}^1\!\!\!\!\mathrm{d}\mu_1\!\!\int_{-1}^1\mathrm{d}\mu_2 (1-\mu_1^2)\nonumber \\
&&(1-\mu_2^2)\left(\frac{1}{1 + \pi i(U\tau/L)k(\mu_2-\mu_1)}\right) \sin^2\left(\frac{\pi}{2}k\mu_1\right)\sin^2\left(\frac{\pi}{2}k\mu_2\right)\nonumber\\ 
&& j_0\left(\frac{\pi}{2}k\mu_1\right) j_0\left(\frac{\pi}{2}k\mu_2\right)\left[\cos\left(\frac{\pi}{2}k\mu_2\right)j_0\left(\frac{\pi}{2}k\mu_1\right) \right.\nonumber\\
&&\left.\qquad\qquad\qquad\qquad\qquad+ \cos\left(\frac{\pi}{2}k\mu_1\right)j_0\left(\frac{\pi}{2}k\mu_2\right)\right],\nonumber\\
\label{covarRTP}
\end{eqnarray}
where $j_0(z) = \sin z/z$ is the spherical Bessel's function of the first kind \citep{im1980table}. The correlated variance is obtained by setting $r=0$ in (\ref{covarRTP}), and is expressible as:
\begin{eqnarray}
\langle\bo{u}(\bo{x})\bo{\cdot}\bo{u}(\bo{x})\rangle\vert_{corr} &=& \frac{3(nL^3)^2}{64\pi^5(\ln\kappa)^3}\left(\frac{U\tau}{L}\right) \int_0^\infty \mathrm{d}k \frac{1}{k^4}\int_{-1}^1\mathrm{d}\mu_1\int_{-1}^1\mathrm{d}\mu_2 (1-\mu_1^2)(1-\mu_2^2)\nonumber \\
&&  \left(\frac{1}{1 + \pi i(U\tau/L)k(\mu_2-\mu_1)}\right)  \sin^2\left(\frac{\pi}{2}k\mu_1\right)\sin^2\left(\frac{\pi}{2}k\mu_2\right) j_0\left(\frac{\pi}{2}k\mu_1\right) \nonumber\\
&&j_0\left(\frac{\pi}{2}k\mu_2\right)\left[\cos\left(\frac{\pi}{2}k\mu_2\right)j_0\left(\frac{\pi}{2}k\mu_1\right)+ \cos\left(\frac{\pi}{2}k\mu_1\right)j_0\left(\frac{\pi}{2}k\mu_2\right)\right].\nonumber\\
\label{varRTP_corr}
\end{eqnarray}
In figure \ref{fig4}, we plot the correlated variance given by (\ref{varRTP_corr}). In accordance with earlier scaling arguments, the correlated variance in suspensions of interacting RTPs takes the form, $(nL^3)^2[c_1 \ln[U\tau/L] + c_2]$ for large $U\tau/L$. In contrast, for $U\tau/L \ll 1$, owing to the more rapid decay of $\Omega_2^{(1)}$, the correlated variance scales as, $c_1 (U\tau/L)(nL^3)^2$ (see inset of figure \ref{fig4}). Next, in figure \ref{fig5}, the correlated covariance is plotted for pusher-type suspensions. For rapid tumblers ($U\tau/L \ll 1$), the covariance directly transitions from the initial variance plateau to an \textit{O}($1/r$) decay for $r\gg 1$. In this limit, one can obtain a simplified form of the correlated covariance from (\ref{covarRTP}), on setting $i(U\tau/L)k(\mu_2-\mu_1)$ to zero, which gives:
\begin{eqnarray}
\langle\bo{u}(\bo{x})\bo{\cdot}\bo{u}(\bo{x} + \bo{r})\rangle\vert_{corr(U\tau/L\to 0)} &=&  \frac{3(nL^3)^2}{128\pi^6(\ln\kappa)^3 r}\left(\frac{U\tau}{L}\right) \int_0^\infty \mathrm{d}k \frac{\sin(2\pi k r)}{k^5}\int_{-1}^1\mathrm{d}\mu_1\!\!\int_{-1}^1\!\!\!\mathrm{d}\mu_2 \nonumber \\
&&(1-\mu_1^2)(1-\mu_2^2) \sin^2\left(\frac{\pi}{2}k\mu_1\right)\sin^2\left(\frac{\pi}{2}k\mu_2\right)j_0\left(\frac{\pi}{2}k\mu_1\right)\nonumber\\ 
&& j_0\left(\frac{\pi}{2}k\mu_2\right) \left[\cos\left(\frac{\pi}{2}k\mu_2\right)j_0\left(\frac{\pi}{2}k\mu_1\right) \right.\nonumber\\
&&\left. \qquad\qquad\quad + \cos\left(\frac{\pi}{2}k\mu_1\right)j_0\left(\frac{\pi}{2}k\mu_2\right)\right].
\label{covarRTP2}
\end{eqnarray}
Setting $r=0$ in (\ref{covarRTP2}), readily yields the correlated variance to be of the form $c_1 (U\tau/L)(nL^3)^2$, consistent with the scaling arguments above.  In contrast, for $U\tau/L \gg 1$, there emerges a weak intermediate regime for $1\ll r\ll U\tau/L$, which delays the onset of the eventual \textit{O}($1/r$) far-field decay (this weak scaling of the covariance in the intermediate region is a logarithmic one, scaling as \textit{O}($\ln[r/(U\tau)]$), and will be derived in \textsection\ref{subsubsec:matched_asymptotic}). One may simplify (\ref{covarRTP}) to obtain the limiting form of this latter far-field dipole asymptote, which may be expressed as: $\langle\bo{u}(\bo{x})\bo{\cdot}\bo{u}(\bo{x} + \bo{r})\rangle\vert_{corr} = 3 (nL^3)^2 (U\tau/L)(28800 \pi r)^{-1}$. The inset in figure \ref{fig5} plots the correlated covariance as a function of $r (U\tau/L)^{-1}$, highlighting the far-field collapse.

\begin{figure}
\includegraphics[width=\textwidth]{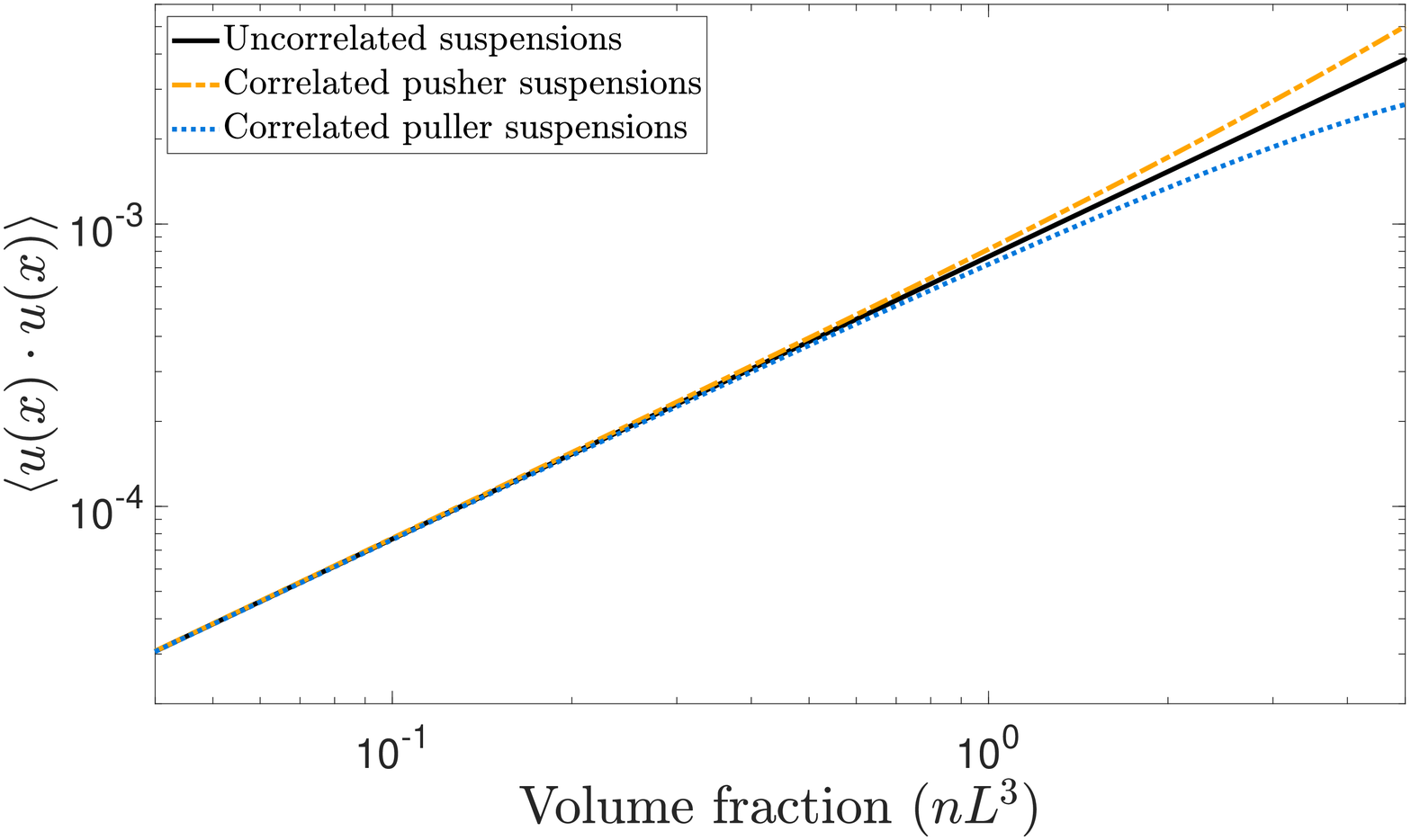}
\caption{\label{fig4a} The total velocity variance plotted as a function of the swimmer volume fraction ($nL^3$) at $U\tau/L = 125$, for both pushers and pullers, along with the uncorrelated contribution (solid line). The swimmer aspect ratio is chosen to be  $\kappa = 8$.}
\end{figure}

One may now write down the expression for the total variance in a suspension of interacting RTPs by adding the variance expression for suspensions of non-interacting swimmers given below (\ref{covarSingleReal}) (see \textsection\ref{subsec:noint}), and the correlated contribution to the variance given by (\ref{varRTP_corr}) above. This is expressible as:
\begin{eqnarray}
\langle\bo{u}(\bo{x})\bo{\cdot}\bo{u}(\bo{x})\rangle =  \frac{1}{2\pi^5}\int_0^\infty \mathrm{d}k &&\frac{1}{k^4}\Bigg\{ \frac{(nL^3)}{(\ln\kappa)^2} \int_{-1}^1\mathrm{d}\mu\left(\frac{1-\mu^2}{\mu^2}\right)\sin^4\left(\frac{\pi k \mu}{2}\right)\nonumber\\
&&+ \frac{3(nL^3)^2}{32(\ln\kappa)^3}\left(\frac{U\tau}{L}\right) \int_{-1}^1\!\!\mathrm{d}\mu_1\int_{-1}^1\mathrm{d}\mu_2 (1-\mu_1^2)(1-\mu_2^2)\nonumber \\
&&\quad\left(\frac{1}{1 + \pi i(U\tau/L)k(\mu_2-\mu_1)}\right) \sin^2\left(\frac{\pi}{2}k\mu_1\right)\sin^2\left(\frac{\pi}{2}k\mu_2\right)\nonumber\\ 
&&\quad  j_0\left(\frac{\pi}{2}k\mu_1\right) j_0\left(\frac{\pi}{2}k\mu_2\right)\left[\cos\left(\frac{\pi}{2}k\mu_2\right)j_0\left(\frac{\pi}{2}k\mu_1\right) \right.\nonumber\\
&&\left.\quad\qquad\qquad\qquad\qquad\qquad+ \cos\left(\frac{\pi}{2}k\mu_1\right)j_0\left(\frac{\pi}{2}k\mu_2\right)\right]\Bigg\}.\nonumber\\
\label{varRTP_Total}
\end{eqnarray}
Expectedly, the total variance scales as $c_1(nL^3) + (nL^3)^2[c_2 \ln[U\tau/L] + c_3]$ for large $U\tau/L$, and as $c_1(nL^3) +c_2 (U\tau/L)(nL^3)^2$ for $U\tau/L\ll 1$. An important implication of including correlations is the scaling of the variance with the dipole strength. The \textit{O}($nL^3$) contribution of the variance has a dimensional scaling of $U^2$, implying that it scales as the square of the dipole strength $D$ (measured in units of $\eta U L^2$). However, the correlated contribution to the variance is proportional to $U^3$, and thence, to $D^3$. Thus, the variance for a suspension of pairwise interacting RTPs depends on the sign of $D$ ($D < 0$ for pushers and $>0$ for pullers), and therefore, on the swimming mechanism. Consistent with recent simulations \citep{deepak2015, stenhammar2017}, at \textit{O}$(nL^3)^2$, (\ref{varRTP_Total}) predicts enhanced fluctuations in pusher suspensions as shown in figure \ref{fig4a}.

\begin{figure}
\includegraphics[width=\textwidth]{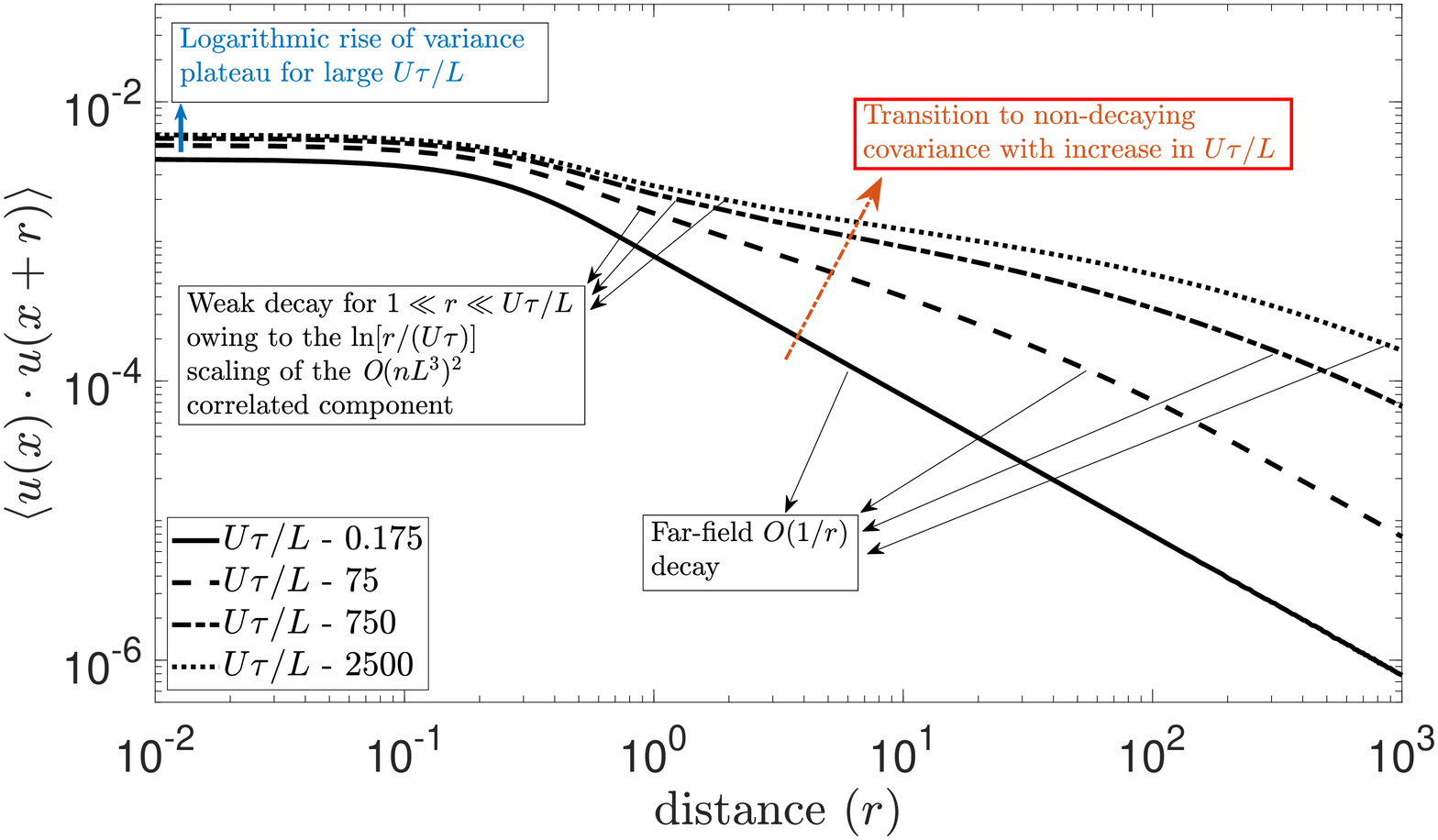}
\caption{\label{fig5a}The total fluid velocity covariance for a suspension of pusher-type RTPs plotted as a function of $r$ over a range of $U\tau/L\in [0.125, 750]$, and for $nL^3 = 5$. The red (dash-dotted) arrow highlights the direction of increasing $U\tau/L$.}
\end{figure}

The total covariance in a suspension of interacting pusher-type RTPs can be similarly expressed by combining (\ref{covarSingleReal}) and (\ref{covarRTP}), which  gives: 
\begin{eqnarray}
\langle\bo{u}(\bo{x})\bo{\cdot}\bo{u}(\bo{x}+\bo{r})\rangle =  \frac{1}{4\pi^6 r}\int_0^\infty && \mathrm{d}k \frac{\sin(2\pi kr)}{k^4}\Bigg\{ \frac{(nL^3)}{(\ln\kappa)^2} \int_{-1}^1\mathrm{d}\mu\left(\frac{1-\mu^2}{\mu^2}\right)\sin^4\left(\frac{\pi k \mu}{2}\right)\nonumber\\
&&+ \frac{3(nL^3)^2}{32(\ln\kappa)^3}\left(\frac{U\tau}{L}\right) \int_{-1}^1\!\!\mathrm{d}\mu_1\int_{-1}^1\mathrm{d}\mu_2 (1-\mu_1^2)(1-\mu_2^2)\nonumber \\
&&\quad\left(\frac{1}{1 + \pi i(U\tau/L)k(\mu_2-\mu_1)}\right) \sin^2\left(\frac{\pi}{2}k\mu_1\right)\sin^2\left(\frac{\pi}{2}k\mu_2\right)\nonumber\\ 
&&\quad  j_0\left(\frac{\pi}{2}k\mu_1\right) j_0\left(\frac{\pi}{2}k\mu_2\right)\left[\cos\left(\frac{\pi}{2}k\mu_2\right)j_0\left(\frac{\pi}{2}k\mu_1\right) \right.\nonumber\\
&&\left.\quad\qquad\qquad\qquad\qquad\qquad+ \cos\left(\frac{\pi}{2}k\mu_1\right)j_0\left(\frac{\pi}{2}k\mu_2\right)\right]\Bigg\}.\nonumber\\
\label{covarRTP_Total}
\end{eqnarray}
Figure \ref{fig5a} shows a plot of the total covariance as given by (\ref{covarRTP_Total}). Note that the chosen $nL^3$ is 5 (and, therefore, a little greater than unity); there is evidence, from earlier calculations for suspensions of slender fibers, that the dilute theory, with pair-interactions accounted for, remains valid for $nL^3$ up until around 7 \citep{mackaplow1996}. In figure \ref{fig5a}, for large $U\tau/L$, while the variance plateau grows slowly with increase in $U\tau/L$, owing to the  $\ln U\tau/L$ scaling of the correlated component, the far-field \textit{O}($1/r$) scaling grows faster on account of the linear scaling with $U\tau/L$ (see the expression for the far-field asymptote below (\ref{covarRTP2})). In addition, for large $U\tau/L$, there is an intermediate regime of shallow decay for separations $1\ll r\ll U\tau/L$, owing to the \textit{O}($\ln[r/(U\tau)]$) scaling of the correlated covariance, which delays the onset of the eventual \textit{O}($1/r$) far-field decay of the covariance. Therefore, as $U\tau/L\to\infty$, one asymptotes to a non-decaying covariance, as expected for the singular straight-swimmer limit discussed earlier.

\subsubsection{Suspensions of interacting RTPs - large $U\tau/L$ covariance using matched asymptotic expansions}\label{subsubsec:matched_asymptotic}

Until now, we have established the detailed expressions for the covariance in suspensions of hydrodynamically interacting straight-swimmers in \textsection\ref{subsubsec:intSt}, and in suspensions of interacting RTPs for arbitrary $U\tau/L$ in \textsection\ref{subsubsec:RTP}. While the limiting form of the covariance in the rapid tumbling limit ($U\tau/L \ll 1$) is readily inferred from the general expression, and is stated in (\ref{covarRTP2}), the form of the covariance in the straight-swimmer limit is more subtle. Owing to a separation of scales between the swimmer size and the run length, one may, however, use the method of matched asymptotic expansions to obtain the covariance in the limit $U\tau/L \gg 1$. In a reference frame moving with one of the swimmers (the test swimmer), the physical domain may be divide into two regions as follows:
\begin{enumerate}
\item An inner region $r\sim$ \textit{O}(1) $\ll U\tau/L$, where the pair of swimmers move relative to each other predominantly in straight trajectories. In this region, one may approximate the correlations induced by the swimmer disturbance velocity fields as those along straight-swimmer trajectories. The expressions pertaining to this region have already been obtained as part of the straight-swimmer analysis given in \textsection\ref{subsubsec:intSt}.
\item An outer region $r\gg U\tau/L$, where orientation decorrelation due to random tumbles need to be taken into account. However, given the large separations, the pair of swimmers may be treated as point force-dipoles. Neglect of the finite swimmer size allows for a considerable simplification of the expressions involved.
\end{enumerate}

\begin{figure}
\centering
\includegraphics[scale=1]{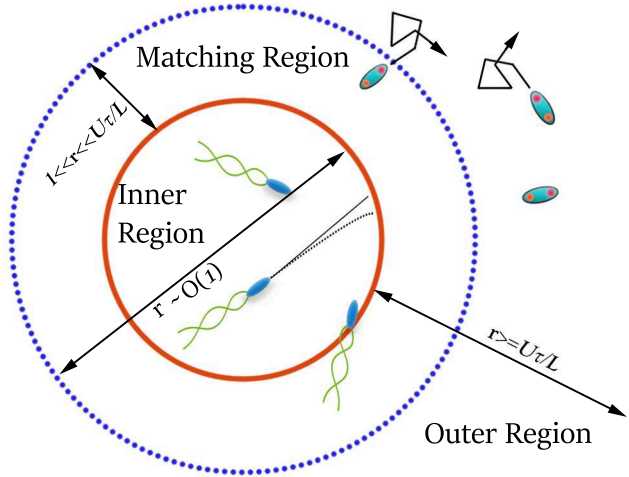}
\caption{A schematic showing the three distinct physical regions for $U\tau/L\gg 1$. The inner region corresponds to $r\sim$ \textit{O}(1), where we have slender straight-swimmers, and the correlations develop over straight trajectories. The outer region $r\sim U\tau/L$ corresponds to the run-and-tumble dynamics of point force-dipoles. In the intermediate region of overlap, $1 \ll r \ll U\tau/L$, the inner and outer solutions must assume a common asymptotic form.}
\label{Schematic}
\end{figure}

A schematic highlighting the above demarcation of the physical domain, for purposes of the matched asymptotic analysis, is shown in figure \ref{Schematic}. One expects that in the inner region the covariance has a non-decaying character on account of the strong correlations between interacting straight-swimmers. In the outer region, there is a transition to the eventual \textit{O}($1/r$) decay for $r \gg U\tau/L$, characteristic of interacting RTP's. The covariance plotted in figure \ref{fig5} does conform to these scalings. However, in the interval $1 \ll r \ll U\tau/L$, corresponding to the region of overlap, there is only a weak decay with separation. The analysis presented in this section enables us to determine the functional form of the covariance in this intermediate region, where the swimmers may be approximated as straight-swimming point force-dipoles.

Now, the point force-dipole, used for modeling the swimmers in the intermediate and outer regions above, has velocity disturbance and pressure fields that satisfy the following equations:
\begin{subeqnarray}
-\bnabla P_i + \eta \nabla^2 \boldsymbol{u}_i(\boldsymbol{x}) &=& D \bo{p}_i\bo{p}_i\bo{:}\bnabla_{\bo{x}}\bo{\delta}(\bo{x}-\bo{x}_i), \label{stokesDP}\\[3pt]
	\bnabla\bo{\cdot}\boldsymbol{u}_i &=& 0. \label{eq:continuityDP}
\end{subeqnarray}
where $D = 1/(4\ln\kappa)$ is the non-dimensional dipole strength and $\bo{p}_i$ denotes the orientation of the $i^{th}$ dipole swimmer. The Fourier transformed velocity covariance for interacting point force-dipoles is then expressible as:
\begin{equation}
\langle\hat{\bo{u}}(\bo{k})\bo{\cdot}\hat{\bo{u}}(\bo{k}^\prime)\rangle\vert_{corr}^{dp} = \frac{D^2}{4\pi^2(\ln\kappa) k^4}\delta(\bo{k}+\bo{k}^\prime) \int \mathrm{d}\bo{p}_1\mathrm{d}\bo{p}_2 (\bo{k}\bo{\cdot}\bo{p}_1)(\bo{k}\bo{\cdot}\bo{p}_2) \left(\mathsfbi{I} - \hat{\bo{k}}\hat{\bo{k}}\right)\bo{:}\bo{p}_1\bo{p}_2\, \hat{\Omega}_{2\,dp}^{(1)}.
\label{eq:fourierDipole}
\end{equation}
where $\hat{\Omega}_{2\,dp}^{(1)}$ denotes the Fourier transformed pair-probability density, and the subscript $dp$ denotes dipoles. The use of the Fourier transformed point-dipole fields obtained from (\ref{stokesDP}), along with the procedure outlined in \textsection\ref{subsubsec:RTP}, leads to the following simplified expression for $\hat{\Omega}_{2\,dp}^{(1)}$ in the outer region:
\begin{equation}
\hat{\Omega}_{2\,dp}^{(1)} = \frac{3D (nL^3)^2}{16\pi^2}\left(\frac{1}{1 + \pi i \frac{U\tau}{L}\bo{k}\bo{\cdot}(\bo{p}_2-\bo{p}_1)}\right) \left(\mathsfbi{I}-\hat{\bo{k}}\hat{\bo{k}}\right)\bo{:}\bo{p}_1\bo{p}_2 (\bo{k}\bo{\cdot}\bo{p}_1)(\bo{k}\bo{\cdot}\bo{p}_2)  .
  \label{eq:omega21FullDipoleT}
\end{equation}

Based on the aforementioned arguments, and with the intent of arriving at an asymptotic expression for large $U\tau/L$, one may partition the original covariance integral in (\ref{eq:fourier}) into inner and outer contributions in the following manner:
\begin{eqnarray}
\langle\bo{u}(\bo{x})\bo{\cdot}\bo{u}(\bo{x}+\bo{r})\rangle\vert_{corr} &=& \frac{2}{r_1}\int_0^{(U\tau/L)^\epsilon} \mathrm{d}k_1 \sin(2\pi k_1 r_1)  \zeta_{dpT}(\bo{k}_1, \bo{p}_1, \bo{p}_2)\nonumber \\ &&\mbox{} + \frac{2}{r}\int_{(U\tau/L)^{(\epsilon-1)}}^\infty \mathrm{d}k \sin(2\pi k r)  \zeta_{SS}(\bo{k}, \bo{p}_1, \bo{p}_2),
\label{eq:matched2}
\end{eqnarray}
where we have introduced a parameter $\epsilon \in (0,1)$ \citep{subramanian2006trajectory}. In (\ref{eq:matched2}), the subscripts $SS$ and $dpT$ denotes slender straight-swimmers and tumbling force-dipole swimmers; here, $\zeta_{SS}(\bo{k}, \bo{p}_1, \bo{p}_2)$ and $\zeta_{dpT}(\bo{k}_1, \bo{p}_1, \bo{p}_2)$, respectively, represent the factors multiplying $\bo{\delta}(\bo{k}+\bo{k}^\prime)$ in (\ref{eq:fourier}) and (\ref{eq:fourierDipole}). Note that, in terms of the wavenumber $k$, the inner and outer regions above correspond to $kL$ and $kU\tau$, respectively, being \textit{O}(1), and this differing non-dimensionalzation is accounted for in (\ref{eq:matched2}). Thus, in the second integral, the scalings are the same as before, with $r$ being the radial distance scaled by $L$; on the other hand, in the first integral $r_1 = r L/(U\tau)$. 

On naively applying the limit $U\tau/L\to \infty$ in (\ref{eq:matched2}), one encounters divergent integrals. This is because $\zeta_{dpT}$ diverges for $k_1\to\infty$ (point dipoles exhibit a divergence as $r_1\to 0$), while $\zeta_{SS}$ diverges for $k\to 0$ (the logarithmic divergence of straight-swimmers as $r\to\infty$). Subtracting the straight-swimming point force dipole contribution (corresponding to the contribution of the overlap region described above) from the respective integrands can, however, account for the said divergences. Therefore, one can recast (\ref{eq:matched2}) as:
\begin{eqnarray}
\langle\bo{u}(\bo{x})\bo{\cdot}\bo{u}(\bo{x}+\bo{r})\rangle\vert_{corr} &=& \frac{2}{r_1}\int_0^{(U\tau/L)^\epsilon} \mathrm{d}k_1 \sin(2\pi k_1 r_1) \Big[ \zeta_{dpT}(\bo{k}_1, \bo{p}_1, \bo{p}_2)\nonumber \\ 
&&\qquad\qquad-  H(k_1-L_o)  \zeta_{dpS}(\bo{k}_1, \bo{p}_1, \bo{p}_2)\Big]\nonumber \\ 
&& \mbox{} + \frac{2}{r}\int_{(U\tau/L)^{(\epsilon-1)}}^\infty \mathrm{d}k \sin(2\pi k r) \Big[ \zeta_{SS}(\bo{k}, \bo{p}_1, \bo{p}_2) \nonumber \\ 
&& \mbox{} \qquad\qquad -  H(k-U_p) \zeta_{dpS}(\bo{k}, \bo{p}_1, \bo{p}_2)\Big]\nonumber \\ 
&& \mbox{} + \frac{2}{r_1}\int_{0}^{(U\tau/L)^\epsilon} \mathrm{d}k_1 \sin(2\pi k_1 r_1)  H(k_1-L_o) \zeta_{dpS}(\bo{k}_1, \bo{p}_1, \bo{p}_2) \nonumber \\ 
&& \mbox{} + \frac{2}{r}\int_{(U\tau/L)^{(\epsilon-1)}}^{\infty} \mathrm{d}k \sin(2\pi k r) H(k-U_p) \zeta_{dpS}(\bo{k}, \bo{p}_1, \bo{p}_2),
\label{eq:matched5}
\end{eqnarray}
such that the divergent contributions are now contained in the final two terms on the right hand side of (\ref{eq:matched5}). Here, the subscript $dpS$ denotes straight-swimming point force-dipoles. Now, the added integral terms are divergent both as $k\to 0$ and as $k\to\infty$, and the Heaviside functions $H(x)$ included in (\ref{eq:matched5}) ensure that these added terms solely account for the divergences already present (in the limit $U\tau/L \to \infty$) in (\ref{eq:matched5}). Note that both $\zeta_{dpS}(\bo{k}, \bo{p}_1, \bo{p}_2)$ and $\zeta_{dpT}(\bo{k}, \bo{p}_1, \bo{p}_2)$ have a common form as $U\tau/L\to\infty$. To see this one may use the Plemlj-Sokhotski formula on the pair-probability density appearing in $\zeta_{dpT}$, as explained in \textsection\ref{subsubsec:RTP} for slender swimmers. Alternatively, one could also determine the  pair-probability in $\zeta_{dpS}$ following the method outlined in \textsection\ref{subsubsec:intSt}, which gives:
\begin{eqnarray}
   \hat{\Omega}_{2dpS}^{(1)} &=& \frac{3D (nL^3)^2}{32\pi^4}\left[\frac{|\bo{p}_2-\bo{p}_1|^2}{\left(\bo{k}\wedge(\bo{p}_2-\bo{p}_1)\right)\bo{\cdot}\left(\bo{k}\wedge(\bo{p}_2-\bo{p}_1)\right)}\right] \left(\bo{k}\bo{\cdot}(\bo{p}_1+\bo{p}_2)\right)^2  \nonumber \\ 
&&\mbox{} \left[\bo{p}_2\bo{\cdot}\bo{p}_1 - \frac{\left(\bo{k}\bo{\cdot}(\bo{p}_1+\bo{p}_2)\right)^2|\bo{p}_2-\bo{p}_1|^2}{4\left(\bo{k}\wedge(\bo{p}_2-\bo{p}_1)\right)\bo{\cdot}\left(\bo{k}\wedge(\bo{p}_2-\bo{p}_1)\right)}\right] \bo{\delta}\left[\bo{k}\bo{\cdot}(\bo{p}_2-\bo{p}_1)\right]\nonumber\\
&&\mbox{} - \frac{3 D i (nL^3)^2}{16\pi^5 k^2 \bo{k}\bo{\cdot}(\bo{p}_2-\bo{p}_1)} \left(\mathsfbi{I}-\hat{\bo{k}}\hat{\bo{k}}\right)\bo{:}\bo{p}_2\bo{p}_1 (\bo{k}\bo{\cdot}\bo{p}_1)(\bo{k}\bo{\cdot}\bo{p}_2).
\label{eq:omega21FullDipoleS}
\end{eqnarray}
The \enquote*{difference integrals} in (\ref{eq:matched5}) that involve the Heaviside functions are now convergent, and one may therefore set $(U\tau/L)^\epsilon = \infty$ and $(U\tau/L)^{(\epsilon-1)}=0$ in (\ref{eq:matched5}) to get:
\begin{eqnarray}
\langle\bo{u}(\bo{x})\bo{\cdot}\bo{u}(\bo{x}+\bo{r})\rangle\vert_{corr} &=& \left[\frac{2}{r_1}\int_0^{\infty} \mathrm{d}k_1 \sin(2\pi k_1 r_1)  \zeta_{dpT}(\bo{k}_1, \bo{p}_1, \bo{p}_2)\right.\nonumber \\ 
&&\mbox{}\left.- \frac{2}{r_1}\int_{L_o}^{\infty} \mathrm{d}k_1 \sin(2\pi k_1 r_1)  \zeta_{dpS}(\bo{k}_1, \bo{p}_1, \bo{p}_2)\right]\nonumber \\
&&\mbox{} + \left[\frac{2}{r}\int_{0}^\infty \mathrm{d}k \sin(2\pi k r)  \zeta_{SS}(\bo{k}, \bo{p}_1, \bo{p}_2) \right.\nonumber \\
&&\mbox{}\left. -\frac{2}{r}\int_{0}^{U_p} \mathrm{d}k \sin(2\pi k r)  \zeta_{dpS}(\bo{k}, \bo{p}_1, \bo{p}_2)\right]\nonumber \\
&&\mbox{} + \frac{2}{r_1}\int_{L_o}^{(U\tau/L)^\epsilon} \mathrm{d}k_1 \sin(2\pi k_1 r_1)  \zeta_{dpS}(\bo{k}_1, \bo{p}_1, \bo{p}_2) \nonumber \\
&&\mbox{} + \frac{2}{r}\int_{(U\tau/L)^{\epsilon-1}}^{U_p} \mathrm{d}k \sin(2\pi k r)  \zeta_{dpS}(\bo{k}, \bo{p}_1, \bo{p}_2),
\label{eq:matched3}
\end{eqnarray}
We designate the first two terms on the right side as $\langle\bo{u}(\bo{x})\bo{\cdot}\bo{u}(\bo{x}+\bo{r})\rangle|_{corr}^{dpT-dpS}$, the next two as $\langle\bo{u}(\bo{x})\bo{\cdot}\bo{u}(\bo{x}+\bo{r})\rangle|_{corr}^{SS-dpS}$ and the last two $U\tau/L$-dependent terms as $\langle\bo{u}(\bo{x})\bo{\cdot}\bo{u}(\bo{x}+\bo{r})\rangle|_{corr}^{Div}$, and evaluate them separately. Note that terms containing $\epsilon$ are now restricted to $\langle\bo{u}(\bo{x})\bo{\cdot}\bo{u}(\bo{x}+\bo{r})\rangle|_{corr}^{Div}$. To obtain a leading order $\epsilon$-independent form from (\ref{eq:matched3}), we first evaluate $\langle\bo{u}(\bo{x})\bo{\cdot}\bo{u}(\bo{x}+\bo{r})\rangle|_{corr}^{Div}$. This will also yield natural choices of $L_o$ and $U_p$ that are most convenient for the analysis.

After some algebra the $U\tau/L$-dependent terms on the right-hand side of (\ref{eq:matched3}) are expressible as:
\begin{eqnarray}
\langle\bo{u}(\bo{x})\bo{\cdot}\bo{u}(\bo{x}+\bo{r})\rangle|_{corr}^{Div} = &&\frac{2(nL^3)^2}{105\pi(\ln\kappa)^3}\left\{ Ci\left[2\pi r_1 L_o \right] - Ci\left[2\pi r_1 \left(\frac{U\tau}{L}\right)^\epsilon \right] \right.\nonumber\\
&&\left. + j_0\left[2\pi r_1 \left(\frac{U\tau}{L}\right)^\epsilon \right] - j_0\left[2\pi r_1 L_o \right] + Ci\left[2\pi r \left(\frac{L}{U\tau}\right)^{1-\epsilon}\right]  \right.\nonumber\\
&&\left. - Ci\left[2\pi r \, U_p\right] +  j_0\left[2\pi r \, U_p \right] - j_0\left[2\pi r \left(\frac{L}{U\tau}\right)^{1-\epsilon}\right]\right\},
\label{eq:divergentTerms}
\end{eqnarray}
where $Ci$ represents the Cosine integral \citep{im1980table}.  We now use the additive method of constructing a uniformly valid approximation of (\ref{eq:divergentTerms}) at leading order in $\epsilon$ \citep{van1964perturbation, subramanian2006trajectory}. This requires approximate forms of (\ref{eq:divergentTerms}) valid, respectively, in the inner, outer and matching regions.

In the inner region, where $r_1\ll 1$ and $r\sim$ \textit{O}(1) one gets:
\begin{eqnarray}
\langle\bo{u}(\bo{x})\bo{\cdot}\bo{u}(\bo{x}+\bo{r})\rangle|_{corr}^{Div - Inner} &=& \frac{2(nL^3)^2}{105\pi}\left\{ 1 + \ln\left(\frac{U\tau U_p}{L L_o}\right) - j_0\left[2\pi r U_p \right]  \right.\nonumber\\ 
&&\mbox{}\left. \qquad\qquad\;+ \int_0^{2\pi r U_p } \frac{\cos(t) -1}{t}\mathrm{d}t\right\},
\label{eq:divergentTerms2}
\end{eqnarray}
where we have used the approximation $Ci(x) \approx j_0(x)$, for large $x$. In the outer region, one has $r_1\sim$ \textit{O}(1) and $r\gg 1$, and at leading order (\ref{eq:divergentTerms}) simplifies to:
\begin{equation}
\langle\bo{u}(\bo{x})\bo{\cdot}\bo{u}(\bo{x}+\bo{r})\rangle|_{corr}^{Div - Outer} = \frac{2(nL^3)^2}{105\pi}\Bigg\{j_0\left[2\pi r_1 L_o \right] - Ci\left[2\pi r_1 L_o\right]\Bigg\}.
\label{eq:divergentTerms3}
\end{equation}
In the matching region, with $r_1\ll 1$ and $r\gg 1$, one gets
\begin{equation}
\langle\bo{u}(\bo{x})\bo{\cdot}\bo{u}(\bo{x}+\bo{r})\rangle|_{corr}^{Div - Matching} = \frac{2(nL^3)^2}{105\pi}\Bigg\{1 - \gamma_E - \ln(2\pi r_1)\Bigg\}.
\label{eq:divergentTerms4}
\end{equation}
Here, $\gamma_E$ is the Euler-Mascheroni constant. From (\ref{eq:divergentTerms2})-(\ref{eq:divergentTerms4}), we find that $L_o = U_p = 1$  is a convenient choice for the analysis, using which, the leading order $\langle\bo{u}(\bo{x})\bo{\cdot}\bo{u}(\bo{x}+\bo{r})\rangle|_{corr}^{Div}$ is expressible as:
\begin{eqnarray}
\langle\bo{u}(\bo{x})\bo{\cdot}\bo{u}(\bo{x}+\bo{r})\rangle|_{corr}^{Div} &=& \frac{2(nL^3)^2}{105\pi}\Bigg\{1 + \ln\left(\frac{U\tau}{L}\right) - j_0\left[2\pi r \right] + \int_0^{2\pi r } \frac{\cos(t) -1}{t}\mathrm{d}t \nonumber \\
&&\mbox{} + j_0\left[2\pi r_1 \right] - Ci\left[2\pi r_1 \right] - \gamma_E - \ln(2\pi r_1) + 1 \Bigg\}.
\label{eq:divergentTerms5}
\end{eqnarray}

\begin{figure}
\centering
\includegraphics[width=\textwidth]{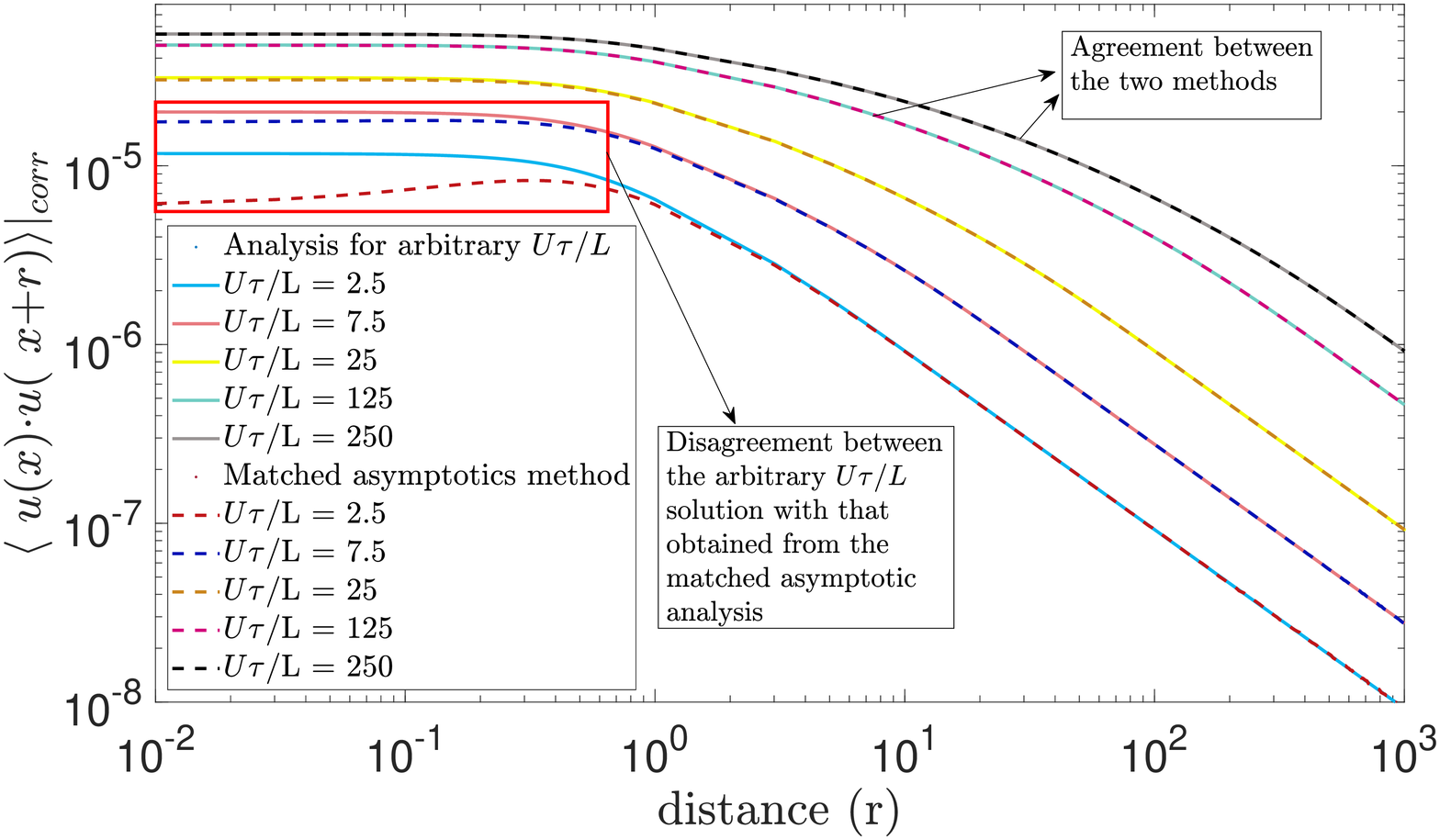}
\caption{The fluid velocity covariance for arbitrary $U\tau/L$ from (\ref{covarRTP}) (straight curves) is plotted alongside the same determined by the method of matched asymptotic expansion (\ref{eq:matched4}) (dashed curves), for $U\tau/L = 2.5, 7.5, 25, 125, 250$.}
\label{matchingPlots}
\end{figure}

\begin{figure}
\centering
\includegraphics[width=\textwidth]{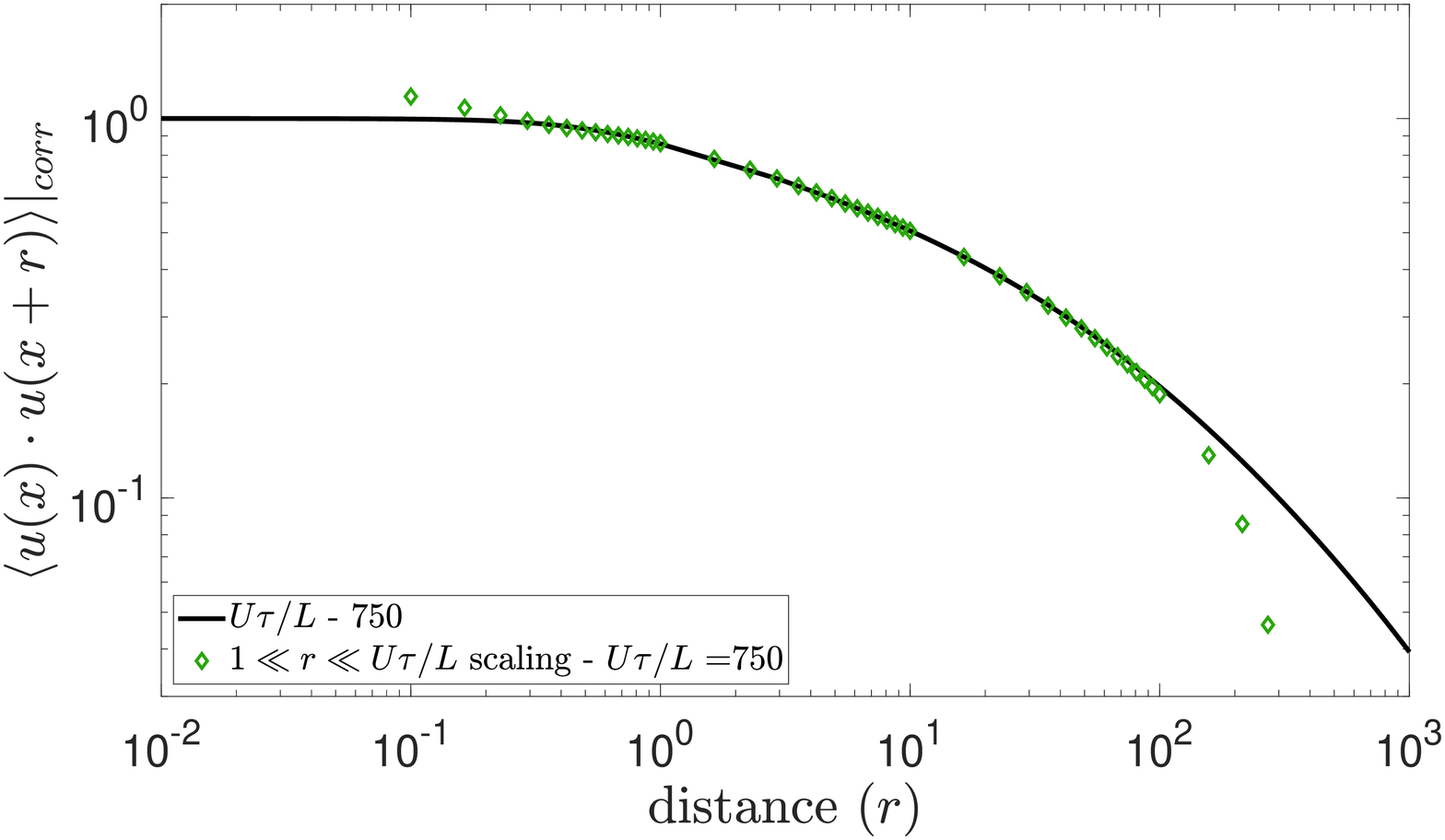}
\caption{The correlated contribution to the fluid velocity covariance for $U\tau/L = 750$  obtained from (\ref{covarRTP}) (straight curves), along with the intermediate  logarithmic scaling ($\diamond$) in the regime $1\ll r\ll U\tau/L$ obtained from (\ref{eq:divergentTerms4}).}
\label{matchingPlots2}
\end{figure}

With $L_o$ and $U_p$ fixed, we now evaluate the remaining terms in (\ref{eq:matched3}). Following standard procedure presented in \textsection\ref{subsubsec:intSt}-\textsection\ref{subsubsec:RTP}, and additionally using (\ref{eq:fourierDipole}), (\ref{eq:omega21FullDipoleT}) and (\ref{eq:omega21FullDipoleS}), $\langle\bo{u}(\bo{x})\bo{\cdot}\bo{u}(\bo{x}+\bo{r})\rangle|_{corr}^{dpT-dpS}$ can be simplified as:
\begin{eqnarray}
\langle\bo{u}(\bo{x})\bo{\cdot}\bo{u}(\bo{x}+\bo{r})\rangle|_{corr}^{dpT-dpS} &=& D^3\Bigg\{\frac{3}{16\pi^2r_1}\int_0^\infty \frac{1}{k_1}\sin(2\pi k_1 r_1)\mathrm{d}k_1\nonumber\\
&&\qquad\quad\,\,\int_{-1}^1\mathrm{d}\mu_1\int_{-1}^1\mathrm{d}\mu_2 \frac{\mu_1^2\mu_2^2(1-\mu_1^2)(1-\mu_2^2)}{1 + \pi^2\left(\frac{U\tau}{L}\right)k_1^2(\mu_2-\mu_1)^2}\nonumber\\ 
&& \mbox{} -\frac{2}{105\pi}\left[j_0(2\pi r_1) - Ci(2\pi r_1)\right]\Bigg\}.
\label{eq:dtds}
\end{eqnarray}
Similarly, using (\ref{eq:stcovar}) for slender straight-swimmers, and (\ref{eq:fourierDipole}) and (\ref{eq:omega21FullDipoleS}) for straight-swimming point force-dipole swimmers, $\langle\bo{u}(\bo{x})\bo{\cdot}\bo{u}(\bo{x}+\bo{r})\rangle|_{corr}^{SS-dpS}$ can be written as:
\begin{eqnarray}
\langle\bo{u}(\bo{x})\bo{\cdot}\bo{u}(\bo{x}+\bo{r})\rangle|_{corr}^{SS-dpS} &=& \frac{D^3}{\pi^2 r}\Bigg\{\int_0^1\mathrm{d}k \int_{-1}^1\mathrm{d}\mu\left[\frac{3}{2\pi^4}\frac{\sin(2\pi k r)}{k^6}(1-\mu^2)^2 \sin(\pi k \mu)\right.\nonumber\\
&&\mbox{}\left.\qquad\qquad\quad \frac{1}{\left(\frac{\pi}{2}k\mu\right)^3}\sin^6\left(\frac{\pi}{2}k\mu\right) - \frac{1}{105}\frac{\sin(2\pi k r)}{k^2}\right]\nonumber\\
&&\mbox{}+\int_1^\infty \mathrm{d}k\frac{\sin(2\pi k r)}{k^6}\int_{-1}^1\mathrm{d}\mu\frac{3}{2\pi^4}(1-\mu^2)^2\frac{1}{\left(\frac{\pi}{2}k\mu\right)^3}\sin^6\left(\frac{\pi}{2}k\mu\right)\Bigg\}.\nonumber\\
\label{eq:ssds}
\end{eqnarray}
Now, using (\ref{eq:divergentTerms5})-(\ref{eq:ssds}) in (\ref{eq:matched3}), we finally get the asymptotic form of $\langle\bo{u}(\bo{x})\bo{\cdot}\bo{u}(\bo{x}+\bo{r})\rangle$ in the limit $U\tau/L\gg 1$ to be:
\begin{eqnarray}
\langle\bo{u}(\bo{x})\bo{\cdot}\bo{u}(\bo{x}+\bo{r})\rangle_{corr} &\equiv & \langle\bo{u}(\bo{x})\bo{\cdot}\bo{u}(\bo{x}+\bo{r})\rangle|_{corr}^{SS-dpS} + \langle\bo{u}(\bo{x})\bo{\cdot}\bo{u}(\bo{x}+\bo{r})\rangle|_{corr}^{dpT-dpS} \nonumber\\
&&\mbox{} + \langle\bo{u}(\bo{x})\bo{\cdot}\bo{u}(\bo{x}+\bo{r})\rangle|_{corr}^{Div}\nonumber \\
&=&\frac{D^3}{\pi^2 r}\Bigg\{\int_0^1\mathrm{d}k \int_{-1}^1\mathrm{d}\mu\left[\frac{3}{2\pi^4}\frac{\sin(2\pi k r)}{k^6}(1-\mu^2)^2 \sin(\pi k \mu)\right.\nonumber\\ &&\left.\qquad\qquad\quad \frac{1}{\left(\frac{\pi}{2}k\mu\right)^3}\sin^6\left(\frac{\pi}{2}k\mu\right) - \frac{1}{105}\frac{\sin(2\pi k r)}{k^2}\right]\nonumber\\
&&\quad\quad\quad+\int_1^\infty \mathrm{d}k\frac{\sin(2\pi k r)}{k^6}\int_{-1}^1\mathrm{d}\mu\frac{3}{2\pi^4}(1-\mu^2)^2\nonumber\\
&&\qquad\qquad\qquad\qquad\qquad\qquad\quad\frac{1}{\left(\frac{\pi}{2}k\mu\right)^3}\sin^6\left(\frac{\pi}{2}k\mu\right)\Bigg\} \nonumber \\
&&+ D^3\Bigg\{\frac{3}{16\pi^2r_1}\int_0^\infty \frac{1}{k_1}\sin(2\pi k_1 r_1)\mathrm{d}k_1\nonumber\\
&&\qquad\quad\,\,\int_{-1}^1\mathrm{d}\mu_1\int_{-1}^1\mathrm{d}\mu_2 \frac{\mu_1^2\mu_2^2(1-\mu_1^2)(1-\mu_2^2)}{1 + \pi^2\left(\frac{U\tau}{L}\right)k_1^2(\mu_2-\mu_1)^2}\nonumber\\
&&\quad\quad\quad -\frac{2}{105\pi}\left[j_0(2\pi r_1) - C_i(2\pi r_1)\right]\Bigg\}\nonumber\\
&& + \frac{2(nL^3)^2}{105\pi}\Bigg\{1 + \ln\left(\frac{U\tau}{L}\right) - j_0\left[2\pi r \right] + \int_0^{2\pi r } \frac{\cos(t) -1}{t}\mathrm{d}t \nonumber \\
&&\quad\quad\quad\quad +  j_0\left[2\pi r_1 \right] - C_i\left[2\pi r_1 \right] - \gamma_E - \ln(2\pi r_1) + 1 \Bigg\}.
\label{eq:matched4}
\end{eqnarray}
It is worth reiterating that while (\ref{covarRTP}) is the exact expression for the velocity covariance valid for arbitrary $U\tau/L$, (\ref{eq:matched4}) is valid only for large $U\tau/L$. In figure \ref{matchingPlots}, we compare the correlated contributions to the covariance calculated from (\ref{covarRTP}) and (\ref{eq:matched4}). The matched asymptotic expansion approach yields excellent agreement with the exact variance, derived in \textsection\ref{subsubsec:RTP} for the three largest values of $U\tau/L$, and remains a good approximation down to $U\tau/L = 7.5$. Up until $U\tau/L\sim$ \textit{O}(1), the covariance smoothly transitions to an \textit{O}($1/r$) far-field scaling following the variance plateau. However, for $U\tau/L\gg 1$, the covariance exhibits a weak intermediate logarithmic decay for $1\ll r\ll U\tau/L$, following which, it again decays as \textit{O}($1/r$) in the far-field. In this regime, $\langle\bo{u}(\bo{x})\cdot\bo{u}(\bo{x}+\bo{r})\rangle|_{corr}^{Div}$ is given by (\ref{eq:divergentTerms4}), and with $r_1 \equiv r*L/(U\tau)$, the weak decay arises from the \textit{O}($\ln [rL/(U\tau)]$) scaling of the \textit{O}$(nL^3)^2$ term in the covariance. The aforementioned weak intermediate logarithmic decay of the covariance is compared to the exact solution, for $U\tau/L = 750$, in figure \ref{matchingPlots2}. 

Therefore, for $U\tau/L\gg 1$, the behavior of the covariance may be summarized as follows:
\begin{enumerate}
\item For $r\ll 1$, it exhibits a finite variance plateau, with the variance itself scaling as $\ln U\tau/L$.
\item For $1\ll r\ll U\tau/L$, it exhibits a weak decay of \textit{O}($\ln [rL/(U\tau)]$).
\item For $r\gg U\tau/L$, corresponding to the far-field, it decays as \textit{O}($1/r$).
\end{enumerate}

\section{Tracer diffusivity}\label{sec:tracers}
The velocity fluctuations analyzed in \textsection\ref{sec:fcovar} convect a passive tracer immersed in the suspension. The dynamics of the passive-tracer is governed by,
\begin{equation}
 \dot{\boldsymbol{x}} = \boldsymbol{u}(\boldsymbol{x}),
\end{equation}
where $\boldsymbol{u}(\boldsymbol{x})$ is the suspension velocity field defined in (\ref{eq:Suspensionoseens}). During a tracer-slender swimmer interaction, the convection velocity of the tracer is \textit{O}$(U/\ln \kappa)$ while the relative tracer-swimmer velocity is \textit{O}$(U)$. Thus, the tracer displacement during an interaction event is negligible and the tracer statistics may be calculated within the Eulerian approximation \citep{kasyapkoch2014}. The expression for the mean-squared displacement (MSD) of the tracer, at time $t$, is then given in terms of the time-dependent fluid velocity correlation function as,
\begin{equation}
\langle x^2(t) \rangle =  \int^t_0 \mathrm{d}t_1 \int^t_0 \mathrm{d}t_2  \langle \boldsymbol{u}(\bo{r}, t_1) \boldsymbol{\cdot}  \boldsymbol{u}(\bo{r}, t_2) \rangle,
\label{eq:msd}
\end{equation}
where the angular-brackets again denote an ensemble average \citep{balakrishnan2008, zwanzig2001}. In terms of the Fourier-transformed variables, the tracer mean-squared displacement is finally given as, 
\begin{equation}
\langle x^2(t) \rangle =  \int \mathrm{d}\bo{k} \int^t_0 \mathrm{d}t_1 \int^t_0 \mathrm{d}t_2  \langle \hat{\boldsymbol{u}}(\bo{k}, t_1) \boldsymbol{\cdot}  \hat{\boldsymbol{u}}(-\bo{k}, t_2) \rangle.
\label{eq:msd}
\end{equation}
Following the discussion in \textsection\ref{sec:fcovar}, we analyze the tracer MSD till \textit{O}($(nL^3)^2$); thus it may be written as,
\begin{equation}
\langle x^2(t) \rangle = \langle x^2(t) \rangle\vert_{uncorr} + \langle x^2(t) \rangle \vert_{corr} ,
\end{equation}
where  $\langle x^2(t) \rangle\vert_{uncorr}  \sim $ \textit{O}($nL^3$) is the tracer MSD in a non-interacting suspension, while $\langle x^2(t) \rangle\vert_{corr}  \sim$ \textit{O}($(nL^3)^2$) denotes the correlated contribution.

\subsection{Suspensions of non-interacting swimmers} \label{subsec:msd_sing}
For non-interacting swimmers, the Fourier-transformed velocity correlation function is given by
\begin{equation}
\langle\hat{\bo{u}}(\bo{k}, t_1)\bo{\cdot} \hat{\bo{u}}(-\bo{k},t_2)\rangle \vert_{uncorr}  =  \int \hat{\boldsymbol{u}}(\boldsymbol{k})\bo{\cdot}\hat{\boldsymbol{u}}(-\boldsymbol{k}) \hat{G}_1 \left(  \bo{p}_1, t_1 \vert \bo{p}_1^\prime, t_2; \bo{k} \right) \Omega_1  \   \mathrm{d}\bo{p}_1 \mathrm{d}\bo{p}_1^\prime,
\end{equation}
where $\Omega_1 = n/4\pi$ is the steady-state singlet probability density and $\hat{G}_1(\bo{p}_1, t_1\vert\bo{p}_1^\prime, t_2; \bo{k})$ is the Fourier transformed transition probability of finding a swimmer with orientation $\bo{p}_1$ at time $t_1$ starting with a swimmer with orientation $\bo{p}_{1}^{\prime}$ at $t_2$. The governing equation for $\hat{G}_1$ is then the same as the equation governing the relaxation of the singlet probability density and is given by,
\begin{equation}
\frac{\mathrm{d} \hat{G}_1}{\mathrm{d} t} + \left(\frac{U\tau}{L}\right)2 \pi i \bo{k} \bo{\cdot} \bo{p}_1 \hat{G}_1 + \left(\hat{G}_1 -\frac{1}{4\pi}\int\mathrm{d}\bo{p}_1 \hat{G}_1 \right) = \bo{\delta}(\bo{p}_1 - \bo{p}_1^\prime)\delta(t - t^\prime) .
\label{eq:G1}
\end{equation}
The orientation-space eigenmodes of the operator governing the transition probability density, $\hat{G}_1$ in (\ref{eq:G1}), that contribute to the velocity-correlation function have no concentration fluctuations (see the discussion in \textsection \ref{subsubsec:RTP}). Thus, for these modes, the inverse-tumbling term in (\ref{eq:G1}) vanishes and the expression for $\hat{G}_1$ is then easily given as, 
\begin{equation}
\hat{G}_1  \left( \bo{p}_1, t \vert \bo{p}_1^\prime, t^{\prime}; \bo{k} \right) = \delta(\bo{p}_1 - \bo{p}_1^\prime) \exp\left(\left(- 2\pi i \bo{k}\bo{\cdot} \bo{p}_1 \frac{U \tau}{L} -1  \right) (t - t^\prime) \right).
\end{equation}
We emphasize that this expression for $\hat{G}_1$ does not capture the relaxation of the concentration modes. The complete expression for $\hat{G}_1$ is needed to characterize the time-dependent evolution of a initially localized population of run-and-tumble swimmers and will be reported separately \citep{ganesh20}. Using the above result, and upon simplifying, we obtain the expression for the MSD of a passive tracer as, 
\begin{eqnarray}
\langle x^2(t) \rangle\vert_{uncorr} & = & \frac{nL^3}{(\ln\kappa)^2 16\pi^7} \int\mathrm{d} \bo{p}_1 \int \frac{\mathrm{d} \bo{k}}{k^4} \left( \frac{-1 + \exp\left(-\left( 2\pi i \bo{k}\bo{\cdot} \bo{p}_1 \frac{U \tau}{L} +1  \right) t \right)}{ \left( 2\pi i \bo{k}\bo{\cdot} \bo{p}_1 \frac{U \tau}{L} +1 \right)^2} \right) \nonumber\\
&& \qquad\qquad\qquad\qquad\quad\left( \mathsfbi{I} - \frac{\bo{k}\bo{k}}{k^2}\right)\bo{:}\bo{p}_1\bo{p}_1\frac{\sin^4\left(\frac{\pi}{2}\bo{k}\bo{\cdot}\bo{p}_1\right)}{(\bo{k}\bo{\cdot}\bo{p}_1)^2} \nonumber\\
&& + \frac{nL^3}{(\ln\kappa)^2 16\pi^7} t  \int \mathrm{d} \bo{p}_1 \int  \frac{\mathrm{d} \bo{k}}{k^4}  \left( \frac{1}{ 2\pi i \bo{k}\bo{\cdot} \bo{p}_1 \frac{U \tau}{L} +1} \right)\left( \mathsfbi{I} - \frac{\bo{k}\bo{k}}{k^2}\right)\bo{:}\!\bo{p}_1\bo{p}_1\nonumber\\
&&\qquad\qquad\qquad\qquad\qquad\quad\frac{\sin^4\left(\frac{\pi}{2}\bo{k}\bo{\cdot}\bo{p}_1\right)}{(\bo{k}\bo{\cdot}\bo{p}_1)^2} .\nonumber\\
\label{eq:msd_sing}
\end{eqnarray}
At times much shorter than the velocity correlation time, the tracer shows ballistic motion and the MSD simplifies to, 
\begin{eqnarray}
\langle x^2(t) \rangle\vert_{uncorr} & = & \frac{t^2}{2} \langle\hat{\bo{u}}(\bo{k}, t)\bo{\cdot} \hat{\bo{u}}(-\bo{k},t)\rangle \vert_{uncorr} \nonumber\\
&& = \frac{nL^3}{(\ln\kappa)^2 32\pi^7} t^2  \int \mathrm{d} \bo{p}_1 \int\frac{\mathrm{d} \bo{k}}{k^4}   \left( \mathsfbi{I} - \frac{\bo{k}\bo{k}}{k^2}\right)\bo{:}\!\bo{p}_1\bo{p}_1\frac{\sin^4\left(\frac{\pi}{2}\bo{k}\bo{\cdot}\bo{p}_1\right)}{(\bo{k}\bo{\cdot}\bo{p}_1)^2}.
\end{eqnarray}
The magnitude of the MSD in the ballistic regime is hence set by the fluid velocity variance discussed in section \textsection\ref{sec:fcovar}. On the other hand at times much longer than the velocity correlation time, the tracer shows diffusive motion with $ \langle x^2(t) \rangle\vert_{uncorr} = D_t\vert_{uncorr} t $, where the effective diffusivity is given by,
\begin{equation}
D_t\vert_{uncorr} =  \frac{nL^3}{(\ln\kappa)^2 16\pi^7}   \int\! \mathrm{d} \bo{p}_1 \int\!\frac{\mathrm{d}  \bo{k}}{k^4}  \left( \frac{1}{ 2\pi i \bo{k}\bo{\cdot} \bo{p}_1 \frac{U \tau}{L} +1} \right)\left( \mathsfbi{I} - \frac{\bo{k}\bo{k}}{k^2}\right)\!\!\bo{:}\!\bo{p}_1\bo{p}_1\frac{\sin^4\left(\frac{\pi}{2}\bo{k}\bo{\cdot}\bo{p}_1\right)}{(\bo{k}\bo{\cdot}\bo{p}_1)^2}  .
\label{eq:msd_sing_diff}
\end{equation}
The diffusivity may also be derived directly using the Green-Kubo formula \citep{balakrishnan2008} and has also been obtained by \cite{kasyapkoch2014} using a different approach. On the other hand, to the best of our knowledge, the complete expression for the MSD in (\ref{eq:msd_sing}) has not been given in the literature even for non-interacting swimmers. As discussed earlier in \textsection \ref{subsec:noint}, the integrals in (\ref{eq:msd_sing}) may be numerically evaluated with the choice of a $\bo{k}$-aligned spherical coordinate system. 

\subsection{Suspensions of interacting swimmers}\label{subsec:msd_int}
\begin{figure}
\includegraphics[scale=0.45]{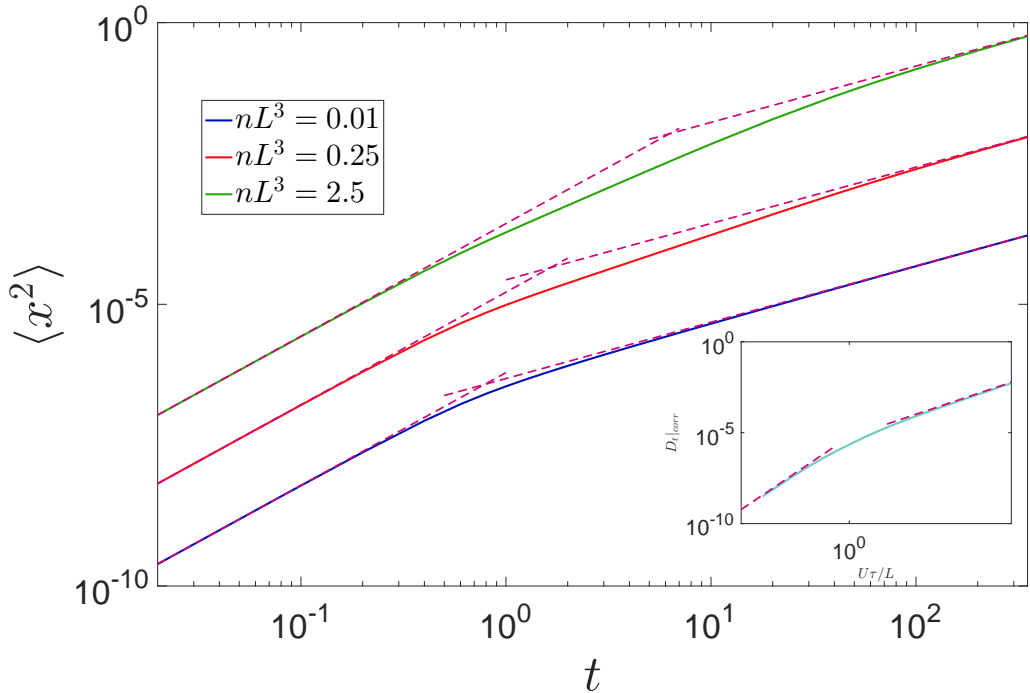}
\caption {The evolution of the mean-squared displacement of the passive tracer, in a suspension of pushers, with time for $U\tau/L =50$ and $\kappa = 8$ for varying $nL^3$. The inset shows the the correlated tracer diffusivity ($D_t\vert_{corr}$) at \textit{O}$(nL^3)^2$ as a function of the run-length, $U\tau/L$, for $\kappa=8$; the dashed lines show the asymptotic scalings discussed in the text.}
\label{fig:diff}
\end{figure}

For interacting swimmers, the Fourier-transformed velocity correlation function at \textit{O}$((nL^3)^2)$ is given by
\begin{eqnarray}
\langle\hat{\bo{u}}(\bo{k}, t_1)\bo{\cdot} \hat{\bo{u}}(-\bo{k},t_2)\rangle \vert_{corr}  =  \int &&\hat{\boldsymbol{u}}(\boldsymbol{k})\bo{\cdot}\hat{\boldsymbol{u}}(-\boldsymbol{k}) \hat{G}_2 \left( \bo{p}_1, \bo{p}_2, t_1 \vert \bo{p}_1^\prime, \bo{p}_2^\prime, t_2; \boldsymbol{k} \right) \hat{\Omega}_2^{(1)} \left( \boldsymbol{k}, \bo{p}_1^\prime, \bo{p}_{2}^{\prime} \right)  \nonumber\\
&& \mathrm{d}\bo{p}_1 \mathrm{d}\bo{p}_1^\prime  \mathrm{d}\bo{p}_2 \mathrm{d}\bo{p}_2^\prime,
\end{eqnarray}
where $\hat{\Omega}_2^{(1)}$ is the correlated pair probability density derived in \textsection\ref{subsubsec:RTP} and $\hat{G}_2$ gives the Fourier transformed  transition probability of finding a pair of swimmers with orientations $\bo{p}_{1}$ and $\bo{p}_{2}$ at time $t_1$ starting with a pair of swimmers with orientations  $\bo{p}_{1}^{\prime}$ and $\bo{p}_{2}^{\prime}$ at time $t_2$. The governing equation for $\hat{G}_2$ is then same as the equation governing the relaxation of the pair probability density and is given by,
\begin{eqnarray}
\frac{\mathrm{d} \hat{G}_2}{\mathrm{d} t}+ 2\pi i \left(\frac{U\tau}{L}\right) \bo{k}\bo{\cdot}(\bo{p}_2-\bo{p}_1)\hat{G}_2 &+& \left(2 \hat{G}_2 -\frac{1}{4\pi}\int\mathrm{d}\bo{p}_1 \hat{G}_2 -\frac{1}{4\pi}\int\mathrm{d}\bo{p}_2 \hat{G}_2 \right) \nonumber\\ 
&&=  \bo{\delta}(\bo{p}_2 - \bo{p}_2^\prime)\bo{\delta}(\bo{p}_1 - \bo{p}_1^\prime)\delta(t - t^\prime).
\label{eq:G2}
\end{eqnarray}
The inverse tumbling terms in (\ref{eq:G2}) again don't contribute (see \textsection\ref{subsubsec:RTP}), and the expression for $\hat{G}_2$ relevant to evaluating the velocity correlation function is given by, 
\begin{equation}
\hat{G}_2  \left( \bo{p}_1, \bo{p}_2, t \vert \bo{p^\prime}_1, \bo{p}_{2}^{\prime}, 0; \boldsymbol{k} \right) = \delta(\bo{p}_1 - \bo{p}_1^\prime) \delta( \bo{p}_2-\bo{p}_{2}^{\prime})  \exp\left(\left(- 2\pi i \bo{k}\bo{\cdot}(\bo{p}_2-\bo{p}_1) \frac{U \tau}{L} -2  \right) t \right).
\end{equation}
Using this result and simplifying, we obtain the final expression for the MSD of a passive tracer at \textit{O}$((nL^3)^2)$ as 
\begin{eqnarray}
\langle x^2(t) \rangle\vert_{corr} & = & \frac{1}{(\ln\kappa)^3 4\pi^6} \int \mathrm{d} \bo{p}_1 \int\mathrm{d} \bo{p}_2 \int \frac{\mathrm{d} \bo{k}}{k^4} \left( \frac{-1 + \exp\left(\left(- 2\pi i \bo{k}\bo{\cdot} (\bo{p}_2 - \bo{p}_1 ) \frac{U \tau}{L} -2  \right) t \right)}{ \left( 2\pi i \bo{k}\bo{\cdot} (\bo{p}_2 - \bo{p}_1) \frac{U \tau}{L} +2 \right)^2} \right)  \nonumber\\ 
&& \qquad\qquad\qquad\left( \mathsfbi{I} - \frac{\bo{k}\bo{k}}{k^2}\right)\bo{:}\bo{p}_1\bo{p}_2\frac{\sin^2\left(\frac{\pi}{2}\bo{k}\bo{\cdot}\bo{p}_1\right)}{(\bo{k}\bo{\cdot}\bo{p}_1)} \frac{\sin^2\left(\frac{\pi}{2}\bo{k}\bo{\cdot}\bo{p}_2\right)}{(\bo{k}\bo{\cdot}\bo{p}_2)} \hat{\Omega}_2^{(1)} \left( \bo{p}_1, \bo{p}_{2}; \boldsymbol{k} \right)  \nonumber\\
&& + \frac{1}{(\ln\kappa)^3 4\pi^6} t  \int \mathrm{d} \bo{p}_1  \int  \mathrm{d} \bo{p}_2  \int \frac{\mathrm{d} \bo{k}}{k^4}  \left( \frac{1}{ 2\pi i \bo{k}\bo{\cdot} (\bo{p}_2 - \bo{p}_1) \frac{U \tau}{L} +2} \right)  \nonumber\\
&& \qquad\qquad\qquad\left( \mathsfbi{I} - \frac{\bo{k}\bo{k}}{k^2}\right)\bo{:}\bo{p}_1\bo{p}_2\frac{\sin^2\left(\frac{\pi}{2}\bo{k}\bo{\cdot}\bo{p}_1\right)}{(\bo{k}\bo{\cdot}\bo{p}_1)} \frac{\sin^2\left(\frac{\pi}{2}\bo{k}\bo{\cdot}\bo{p}_2\right)}{(\bo{k}\bo{\cdot}\bo{p}_2)} \hat{\Omega}_2^{(1)} \left( \bo{p}_1, \bo{p}_{2}; \boldsymbol{k} \right). \nonumber\\
\label{eq:msd_pair}
\end{eqnarray}
At times much shorter than the velocity correlation time, the tracer shows ballistic motion and the MSD simplifies to,
\begin{eqnarray}
\langle x^2(t) \rangle_{corr} & = & \frac{t^2}{2} \langle\hat{\bo{u}}(\bo{k}, t)\bo{\cdot} \hat{\bo{u}}(-\bo{k},t)\rangle \vert_{corr} \nonumber\\ 
&& = \frac{t^2}{(\ln\kappa)^3 8\pi^6}  \int \mathrm{d} \bo{p}_1 \int \mathrm{d} \bo{p}_2 \int\frac{\mathrm{d} \bo{k}}{k^4}  \left( \mathsfbi{I} - \frac{\bo{k}\bo{k}}{k^2}\right)\bo{:}\bo{p}_1\bo{p}_2 \nonumber\\
&&\qquad\qquad\qquad \frac{\sin^2\left(\frac{\pi}{2}\bo{k}\bo{\cdot}\bo{p}_1\right)}{(\bo{k}\bo{\cdot}\bo{p}_1)} \frac{\sin^2\left(\frac{\pi}{2}\bo{k}\bo{\cdot}\bo{p}_2\right)}{(\bo{k}\bo{\cdot}\bo{p}_2)} \hat{\Omega}_2^{(1)} \left( \boldsymbol{k}, \bo{p}_1, \bo{p}_{2} \right)   .
\end{eqnarray}
The magnitude of the MSD in the ballistic regime is again set by the fluid velocity variance, with the relevant $U\tau/L$ scalings discussed in section \textsection\ref{sec:fcovar}. On the other hand, for times much longer than the velocity correlation time, the tracer shows diffusive motion with $ \langle x^2(t) \rangle\vert_{corr} = D_t\vert_{corr} t $ for any finite swimmer run-length ($U\tau/L$). The effective diffusivity is given by,
\begin{eqnarray}
D_t\vert_{corr} & = &  \frac{1}{(\ln\kappa)^3 4\pi^6} \int \mathrm{d} \bo{p}_1 \int\mathrm{d} \bo{p}_2 \int\frac{\mathrm{d}  \bo{k}}{k^4}  \left( \frac{1}{ 2\pi i \bo{k}\bo{\cdot} (\bo{p}_2 - \bo{p}_1) \frac{U \tau}{L} +2} \right)\left( \mathsfbi{I} - \frac{\bo{k}\bo{k}}{k^2}\right)\bo{:}\!\bo{p}_1\bo{p}_2 \nonumber\\
&&\qquad\qquad\qquad \frac{\sin^2\left(\frac{\pi}{2}\bo{k}\bo{\cdot}\bo{p}_1\right)}{(\bo{k}\bo{\cdot}\bo{p}_1)} \frac{\sin^2\left(\frac{\pi}{2}\bo{k}\bo{\cdot}\bo{p}_2\right)}{(\bo{k}\bo{\cdot}\bo{p}_2)} \hat{\Omega}_2^{(1)} \left( \boldsymbol{k}, \bo{p}_1, \bo{p}_{2} \right) .
\label{eq:msd_pair_diff}
\end{eqnarray}
The integrals in (\ref{eq:msd_pair}) and (\ref{eq:msd_pair_diff}) are again evaluated numerically (see \textsection \ref{subsec:noint} for the coordinate system used). The total mean-squared displacement of the tracer, to \textit{O}$(nL^3)^2$, is obtained by combining the expressions in (\ref{eq:msd_sing}) and (\ref{eq:msd_pair}), and is plotted in figure \ref{fig:diff}, as a function of $nL^3$, for a suspension of pushers.

Surprisingly, figure \ref{fig:diff} shows that the time taken to transition from the ballistic to the diffusive regime increases with increasing volume-fraction of the swimmers. This surprising behavior can be explained by noting the differing time scales for the decay of the velocity correlations at \textit{O}$(nL^3)$ and \textit{O}$(nL^3)^2$. At \textit{O}$(nL^3)$, this time-scale ($t_c$) is set by the swimmer-tracer interaction time. For rapid tumblers ($U\tau/L \ll 1$), the interaction is cut off by the decorrelation of the swimmer orientation, implying $t_c \sim $ \textit{O}($\tau$). On the other hand for straight swimmers ($U\tau/L \gg 1$), the distance that the tracer is convected asymptotes to a finite value for tracer-swimmer separations greater than \textit{O}$(L)$, and thus $t_c \sim$ \textit{O}($L/U$). At \textit{O}$(nL^3)^2$, the decay of the velocity correlations, irrespective of $U\tau/L$, only occurs due to orientation decorrelation of the swimmers on the time scale $\tau$. As a result for $nL^3$ of order unity, the transition time between the ballistic and diffusive regimes diverges as $U\tau/L$. The resulting broad cross-over gives the impression of an $nL^3$-dependent anomalous exponent in the interval $L/U  \ll t  \ll \tau$ (see the curve for $nL^3 = 2.5$ in figure \ref{fig:diff}). For rapid tumblers $(U\tau/L \ll 1)$, the transition time is \textit{O}$(\tau)$ regardless of $nL^3$, and there is no intermediate anomalous scaling.

\begin{figure}
\includegraphics[scale=0.45]{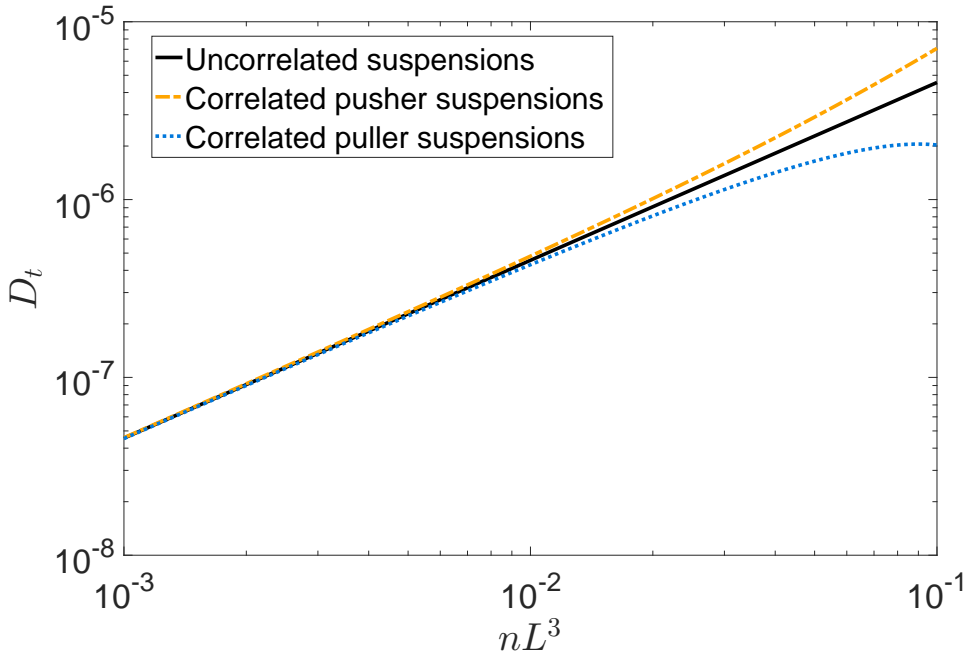}
\caption {The variation of the diffusivity of a passive tracer versus the volume fraction of swimmers ($nL^3$) for a fixed run-length $U\tau/L =50$ and aspect-ratio $\kappa = 8$.}
\label{fig:diffbif}
\end{figure}

The inset of figure \ref{fig:diff} shows the \textit{O}$(nL^3)^2$ correlated contribution to the tracer diffusivity ($D_t\vert_{corr}$). In the straight-swimmer limit, $D_t\vert_{corr}$ diverges linearly in $U\tau/L$. This scaling may be rationalized by starting from $D_t \sim U_{t}^{2} t_c$ where $U_t$ is the scale of the velocity convecting the tracer and $t_c$ is the time scale of decay of the velocity correlations. $U_t$ is \textit{O}$(U/\ln \kappa)$ regardless of the swimmer volume-fraction $(nL^3)$, while the $t_c$ scaling is discussed above. The resulting scaling for tracer diffusivity in an interacting suspension, to \textit{O}$(nL^3)^2$, is therefore given by,
\begin{equation}
\frac{nL^3 U^2 \tau}{(\ln \kappa)^2} (\tilde{d_1} + \frac{nL^3}{\ln \kappa} \frac{U\tau}{L} \tilde{d_2})
   \quad\mathrm{for}\quad 
U\tau/L \ll 1,
\end{equation}
and
\begin{equation}
\frac{nL^4U}{(\ln \kappa)^2} (d_1 + \frac{nL^3}{\ln \kappa} \frac{U\tau}{L} d_2 )
   \quad\mathrm{for}\quad 
U\tau/L \gg 1.
\end{equation}
and confirms the scalings shown in figure \ref{fig:diff} (inset) at \textit{O}$(nL^3)^2$.  The tracer diffusivity variation with volume fraction also shows a pusher-puller bifurcation similar to the velocity variance as shown in figure \ref{fig:diffbif}.

Several experiments probing bacterial suspension dynamics with passive tracers have reported an intermediate super-diffusive regime and a volume fraction dependent cross-over time \citep{Wu2000, kim04,  arratia2016, marenduzzo2011, cheng2016, lubensky2007, volpe2016}. To the best of our knowledge, this is the first theoretical explanation for these observations. Our analysis thus shows that in a bath of interacting persistent swimmers ($U\tau/L \gg 1$), the temporal and spatial correlations of the (effective) noise experienced by the passive tracer increase to \textit{O}$(\tau)$ and \textit{O}$(U\tau)$ respectively. The simplistic approximation for the active noise often found in the literature is thus seen to be insufficient \citep{maggi2014, volpe2016, kanazawa2019}. Further, when confined by a harmonic potential with a trap radius of \textit{O}$(U\tau)$, the probability distribution of the tracer displacements is thus expected to become pronouncedly non-Gaussian with increasing $nL^3$. This has indeed been seen in experiments \citep{sood2016, volpe2016}. 

\section{Conclusion}\label{sec:conclusion}

In this paper, we have shown that long-ranged pair-correlations lead to divergent velocity fluctuations and tracer diffusivities in straight-swimmer suspensions. For RTPs, this divergence is replaced by a logarithmic increase of the velocity variance, and a linear increase of the tracer diffusivity, with the run length for $U\tau/L \gg 1$. Correlated fluctuations are also crucially dependent on the swimming mechanism (pusher vis-a-vis puller), being larger for pusher suspensions. Thus, in contrast to recent work emphasizing the role of anisotropic tracers \citep{cheng2016,cheng2016b}, spherical tracers already discriminate between pusher and puller suspensions due to correlated fluctuations. 

While our detailed analysis is limited to slender swimmers, the scaling arguments presented only involve far-field hydrodynamics. It is thus expected that they would be applicable to suspensions of general anisotropic swimmers \citep{balmforth2014, ishikawa2015} provided far-field interactions dominate over near-field ones. This assertion is supported by the evidence of logarithmic scaling for the velocity variance in simulations of regularized point dipoles by \cite{underhill2011}. The spherical squirmer model often used in studies of active suspensions may therefore represent an exceptional scenario and the results obtained need to be extrapolated with care \citep{ishikawa2006hydrodynamic, ishikawa2007rheology, pedley08, nott08, pagonabarraga2013, climent2018}.  

In this paper, we have examined correlations due to direct pairwise hydrodynamic interactions between swimmers, while not accounting for those induced (indirectly) by long-wavelength fluctuations in the suspension velocity and orientation fields. It is well-known that these fluctuations lead to an instability of a pusher suspension, in turn implying diverging contributions of the \enquote*{collective effects} to statistical correlations at the threshold \citep{subkoch2009, stenhammar2017, underhill17, morozov2019}. This divergence is distinct from the one described here. Nevertheless, the framework needed for this calculation requires the $U\tau/L$-dependent pair-probability, determined here, as an input, and will be reported in future-work. Our work thus lays a rigorous foundation for understanding hydrodynamic interactions between anisotropic swimmers, and their role in bacterial turbulence \citep{yoemans2012}.   

\appendix

\section{Rotation rates of slender swimmers}\label{sec:appRotationRate}

Herein, we derive the expression for the rates of rotation $\dot{\bo{p}}_{ij}$ for a pair of interacting slender swimmers. One can define the rate of rotation of swimmer $1$ due to swimmer $2$ as $\dot{\bo{p}}_{12} = \bo{\omega}_1\wedge\bo{p}_1$, with $\boldsymbol{\omega}_1$ representing its angular velocity. The velocity at any point $s$ on a rigid slender body may be written as $\bo{u}_1^s(s\bo{p}_1) = \bo{u}_1^T + \bo{\omega}_1\wedge\bo{p}_1s$, where $\bo{u}_1^T$ is the translational velocity of the geometric center ($s = 0$). From viscous slender body theory \citep{batchelor1970slender, kim1992microhydrodynamics}, one has the relation:  $\bo{u}_1^s(s\bo{p}_1) -\bo{u}^\infty(s\bo{p}_1) = (\ln\kappa/\eta)f(s)\bo{p}_1\bo{\cdot}(\mathsfbi{I}+\bo{p}_1\bo{p}_1)$ to leading logarithmic order. Here, $\bo{u}^\infty$ is the ambient field which in the present problem is the velocity field generated by the swimmer $2$, that is, $\bo{u}^\infty\equiv\bo{u}_2$; the force density $f(s) = \eta U sgn(s)/\ln\kappa$. To determine, $\bo{u}_2$, one solves the Stokes equations (\ref{stokes}) at any location ($-\bo{r}+s\bo{p}_1$) along swimmer 1; that is:
\begin{equation}
-\nabla P_2 + \eta \nabla_r^2\bo{u}_2(-\bo{r}+s\bo{p}_1) = \int_{-L/2}^{L/2}\mathrm{d}s^\prime f(s^\prime) \bo{\delta}(-\bo{r}+s\bo{p}_1-s^\prime\bo{p}_2)\bo{p}_2.
\label{appendixStokes}
\end{equation}
Again, it is convenient to solve (\ref{appendixStokes}) in Fourier space, and the Fourier transformed ambient field along the length of swimmer $1$, $\hat{u}_2(\bo{k})$, can be written as:
\begin{equation}
\hat{\bo{u}}_2(\bo{k}) = -\frac{i}{2\pi^3k^2}\left[\mathsfbi{I}-\bo{\bo{k}\bo{k}}\right]\bo{\cdot}\bo{p}_2 \exp(2\pi i s \bo{k}\cdot\bo{p}_1)\frac{1}{(\bo{k}\bo{\cdot}\bo{p}_2)}\sin\left(\frac{\pi}{2}\bo{k}\bo{\cdot}\bo{p}_2\right),
\label{disturbanceVelocity}
\end{equation}
on applying the non-dimensionalizations stated below (\ref{covarFourier3}). The force-free condition ensures that the swimmer propagates with the sum of the swimming and disturbance velocities averaged over its length. However, given that the latter is \textit{O}($1/\ln\kappa$) smaller, it may be neglected. Applying the torque-free condition, $\int_{-1}^1 s\bo{p}_1\wedge [\bo{u}_1^s(s\bo{p}_1) -\bo{u}_2(s\bo{p}_1)]\mathrm{d}s = 0$, results in the following expression for $\hat{\dot{\bo{p}}}_{12}$:
\begin{equation}
\hat{\dot{\bo{p}}}_{12} = \frac{3}{\pi^3 k^2}\frac{1}{(\bo{k}\bo{\cdot}\bo{p}_2)}\sin^2\left(\frac{\pi}{2}\bo{k}\bo{\cdot}\bo{p}_2\right) j_1(\pi \bo{k}\bo{\cdot}\bo{p}_1)\left\{\left[\bo{p}_1\wedge(\mathsfbi{I}-\hat{\bo{k}}\hat{\bo{k}})\bo{\cdot}\bo{p}_2\right]\wedge\bo{p}_1\right\},
\label{pdot}
\end{equation}
where $j_1(x) = \sin(z)/z^2 - \cos(z)/z$ is the spherical Bessel's function of the first kind \citep{im1980table}. The Fourier transformed rate of rotation of the second swimmer $\hat{\dot{\bo{p}}}_{21}$ is given by interchanging the indices $1$ and $2$ in (\ref{pdot}).

\section{Orientation Correlations}\label{sec:orientation}

\begin{figure}
\includegraphics[scale=0.625]{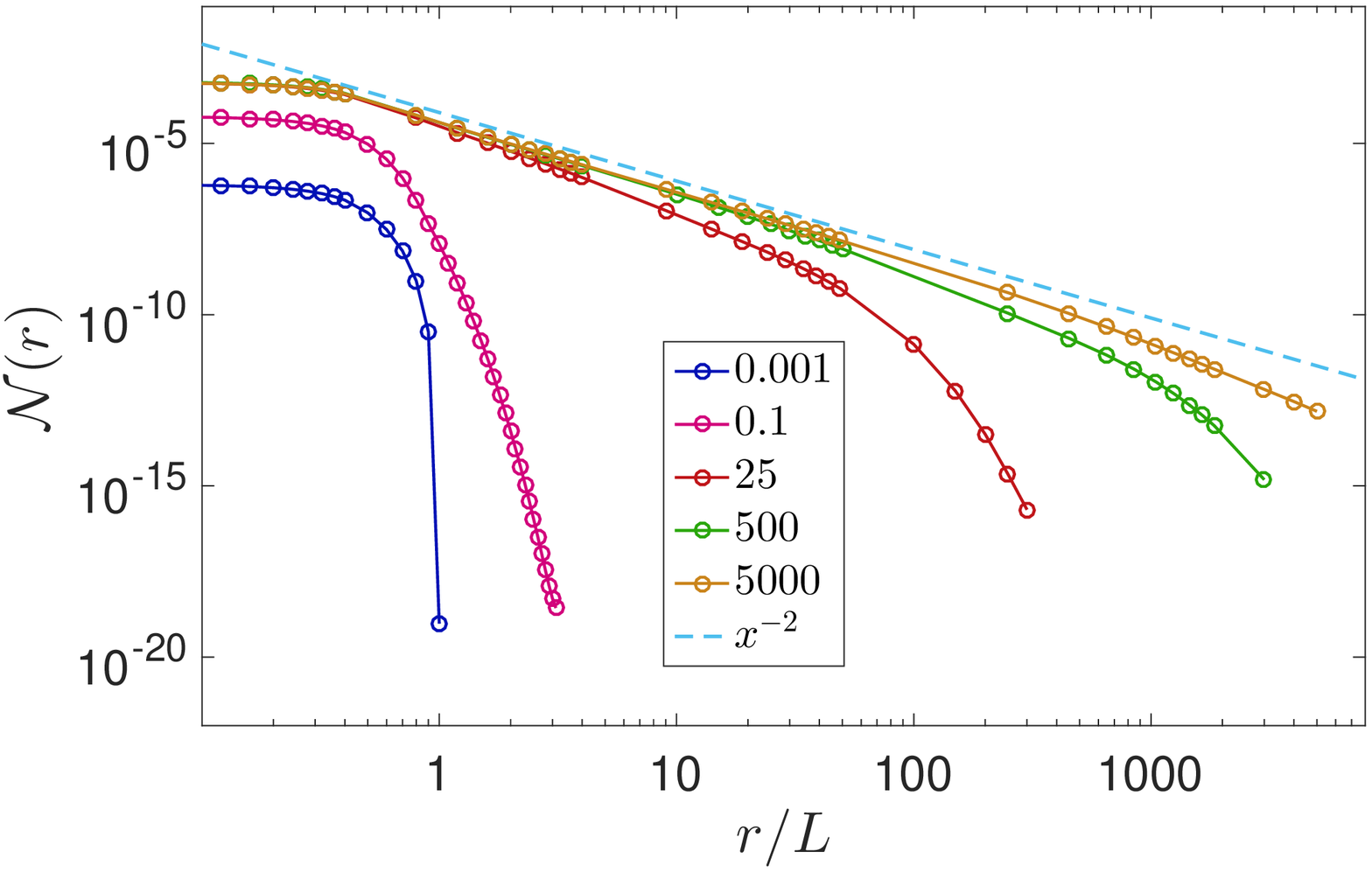}
\caption {The spatial decay of the nematic correlations at \textit{O}$(1/ \ln\kappa)$ for various dimensionless run-lengths $U\tau/L$, in an interacting pusher suspension; the dashed line shows the asymptotic power law discussed in the text.}
\label{fig:nematic}
\end{figure}

\begin{figure}
\includegraphics[scale=0.625]{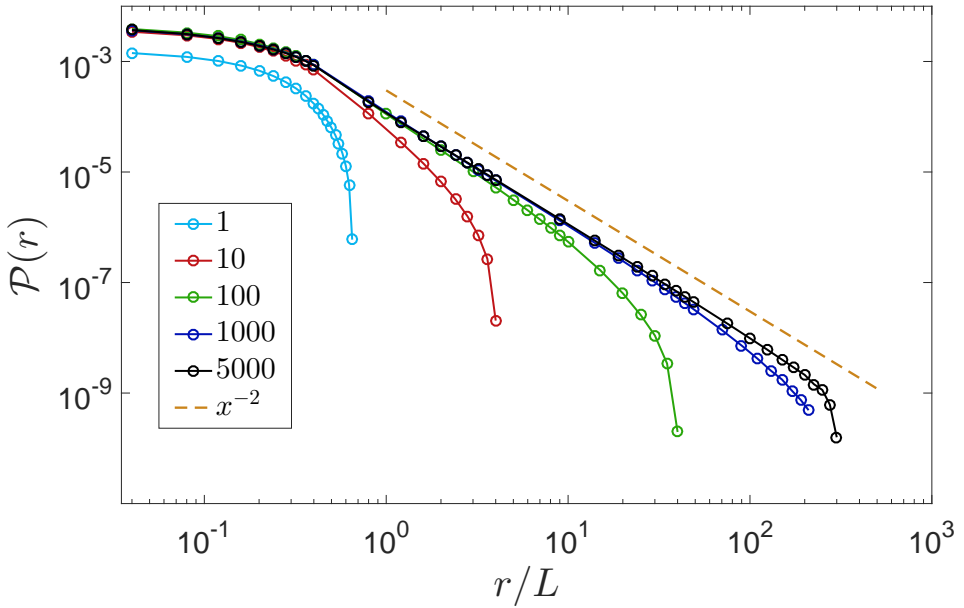}
\caption {The spatial decay of the polar correlations at \textit{O}$(1/ \ln\kappa)$ for various dimensionless run-lengths $U\tau/L$, in an interacting pusher suspension; the dashed line shows the asymptotic power law discussed in the text.}
\label{fig:polar}
\end{figure}

Herein, we interpret the orientation correlations that develop in a suspension of pairwise interacting RTPs, by characterizing the nematic and polar correlations. In  a suspension of interacting swimmers, the nematic correlation, at the leading order, between two swimmers at a distance $r$ is given by,
\begin{eqnarray}
\mathcal{N}(r) &&= \frac{ \langle (\bo{p}_1 \bo{\cdot} \bo{p}_2)^2 \rangle} { \langle 1 \rangle} \nonumber\\
&& = \frac{\int \mathrm{d} \bo{p}_1 \int \mathrm{d} \bo{p}_2  (\bo{p}_1 \bo{\cdot} \bo{p}_2)^2 \Omega_2(\bo{r}, \bo{p}_2, \bo{p}_1)} {\int \mathrm{d} \bo{p}_1 \int \mathrm{d} \bo{p}_2  \Omega_2(\bo{r},\bo{p}_2, \bo{p}_1)} \nonumber\\
&& = \frac{ \int \mathrm{d} \bo{p}_1 \int \mathrm{d} \bo{p}_2  (\bo{p}_1 \bo{\cdot} \bo{p}_2)^2 \hat{\Omega}_2^{(1)}(\bo{r},\bo{p}_2, \bo{p}_1) \exp(2 \pi i \bo{k} \bo{\cdot} \bo{r})} {(\ln\kappa)(nL^3)^2},
\label{eq:nematic_1}
\end{eqnarray} 
whereas, the polar correlation, at the leading order, is expressible as,
\begin{eqnarray}
\mathcal{P}(r) &&=  \frac{ \langle \bo{p}_1 \bo{\cdot} \bo{p}_2 \rangle} { \langle 1 \rangle} \nonumber \\
&& = \frac{\int \mathrm{d} \bo{p}_1 \int \mathrm{d} \bo{p}_2  (\bo{p}_1 \bo{\cdot} \bo{p}_2) \Omega_2(\bo{r}, \bo{p}_2, \bo{p}_1)} {\int \mathrm{d} \bo{p}_1 \int \mathrm{d} \bo{p}_2  \Omega_2(\bo{r},\bo{p}_2, \bo{p}_1)} \nonumber\\
&& = \frac{ \int \mathrm{d} \bo{p}_1 \int \mathrm{d} \bo{p}_2  (\bo{p}_1 \bo{\cdot} \bo{p}_2) \hat{\Omega}_2^{(1)}(\bo{r},\bo{p}_2, \bo{p}_1) \exp(2 \pi i \bo{k} \bo{\cdot} \bo{r})} {(\ln\kappa)(nL^3)^2},
\label{eq:polar_1}
\end{eqnarray} 
where we have used the normalisation condition 
\begin{equation}
\int \mathrm{d} \bo{p}_1 \int \mathrm{d} \bo{p}_2  \Omega_2(\bo{r},\bo{p}_2, \bo{p}_1) =\int \mathrm{d} \bo{p}_1 \int \mathrm{d} \bo{p}_2  \Omega_2^{(0)}(\bo{r},\bo{p}_2, \bo{p}_1) = (nL^3)^2.
\end{equation}
Using the result for $\hat{\Omega}^{(1)}_2$ from (\ref{eq:omega21RTP5}), (\ref{eq:nematic_1}) and (\ref{eq:polar_1}) simplify to,
\begin{eqnarray}
   \mathcal{N}(r) &=&  \frac{3 }{(\ln\kappa) 32\pi^6} \left(\frac{U\tau}{L}\right) \int  \mathrm{d} \bo{k}\frac{1}{k^2} \int \mathrm{d} \bo{p}_1 \int \mathrm{d} \bo{p}_2 \left(\frac{ (\bo{p}_1 \bo{\cdot} \bo{p}_2)^2 }{\pi i (U\tau/L)\bo{k}\bo{\cdot}(\bo{p}_2-\bo{p}_1) + 1}\right) \nonumber\\
 &&\left[\frac{1}{\left(\bo{k}\bo{\cdot}\bo{p}_1\right)} \sin^2\left(\frac{\pi}{2}\bo{k}\bo{\cdot}\bo{p}_1\right)\sin\left(\pi \bo{k}\bo{\cdot}\bo{p}_2\right)
 +  \frac{1}{\left(\bo{k}\bo{\cdot}\bo{p}_2\right)} \sin^2\left(\frac{\pi}{2}\bo{k}\bo{\cdot}\bo{p}_2\right)\sin\left(\pi \bo{k}\bo{\cdot}\bo{p}_1\right) \right]\nonumber\\
 && \left(\mathsfbi{I}-\hat{\bo{k}}\hat{\bo{k}}\right)\bo{:}\bo{p}_2\bo{p}_1  \exp(2 \pi i \bo{k} \bo{\cdot} \bo{r}) \nonumber\\
\end{eqnarray}
and
\begin{eqnarray}
   \mathcal{P}(r) &=&  \frac{3}{(\ln\kappa) 32\pi^6} \left(\frac{U\tau}{L}\right) \int  \frac{\mathrm{d} \bo{k}}{k^2} \int \mathrm{d} \bo{p}_1 \int \mathrm{d} \bo{p}_2  \left(\frac{\bo{p}_1 \bo{\cdot} \bo{p}_2 }{\pi i (U\tau/L)\bo{k}\bo{\cdot}(\bo{p}_2-\bo{p}_1) + 1}\right)\nonumber\\
 &&  \left[\frac{1}{\left(\bo{k}\bo{\cdot}\bo{p}_1\right)} \sin^2\left(\frac{\pi}{2}\bo{k}\bo{\cdot}\bo{p}_1\right)\sin\left(\pi \bo{k}\bo{\cdot}\bo{p}_2\right) +  \frac{1}{\left(\bo{k}\bo{\cdot}\bo{p}_2\right)} \sin^2\left(\frac{\pi}{2}\bo{k}\bo{\cdot}\bo{p}_2\right)\sin\left(\pi \bo{k}\bo{\cdot}\bo{p}_1\right) \right] \nonumber\\
 && \left(\mathsfbi{I}-\hat{\bo{k}}\hat{\bo{k}}\right)\bo{:}\bo{p}_2\bo{p}_1 \exp(2 \pi i \bo{k} \bo{\cdot} \bo{r}),
\end{eqnarray}
respectively. These expressions are numerically evaluated using the coordinate system described in detail in \textsection \ref{subsec:int}. Figures \ref{fig:nematic} and \ref{fig:polar} show the spatial decay of the nematic and polar correlations, respectively, for a range of $U\tau/L$ in a suspension of pushers. The orientation correlations are qualitatively similar to those obtained in  previous numerical computations \citep{saintillan07, morozov2019}. 

For small $U\tau/L$, the orientational correlations are constant for a swimmer length ($r \sim 1$). However, for large distances ($r \gg 1$), each swimmer sees the other swimmer as a rapidly tumbling dipole and hence the correlations decay to zero faster than any power law. On the other hand, in the limit of straight-swimmers ($U\tau/L \gg 1$), there is an intermediate region ($1 \ll r \ll U\tau/L$) where the correlations decay as a power law ($ \mathcal{P}, \mathcal{N}\sim 1/r^2$). For distances $r \gg U\tau/L$, the correlations are again screened due to tumbling. Further, we see that the orientational correlations, for any $r$, asymptote to a constant value for straight swimmers ($U\tau/L \gg 1$) unlike the velocity variance. For rapid tumblers, on the other hand, as expected the nematic correlation, $\mathcal{N}(r)/(U\tau/L) \sim$ \textit{O}$(1)$, dominates over the polar correlation, $\mathcal{P}(r)/(U\tau/L) \sim$ \textit{O}$((U \tau /L)^2)$.

\bibliographystyle{jfm}
\bibliography{refer}

\end{document}